\documentclass[12pt]{article}
\usepackage{amssymb}
\usepackage{amsmath}
\usepackage{amsthm}
\usepackage{cancel}
\usepackage{slashed}
\usepackage{fullpage}
\usepackage{color}
\usepackage{braket}
\usepackage{graphicx}
\usepackage[small]{subfigure}
\usepackage{multirow}
\usepackage{verbatim}
\usepackage{cite}
\usepackage{bigints}
\usepackage{simplewick}
\usepackage{authblk}
\usepackage{hyperref}
\usepackage{empheq}
\usepackage{tikz}

\hypersetup{
    linktocpage,
     colorlinks,
     citecolor=darkgreen,
     linkcolor= darkgreen,
     urlcolor=darkgreen
}

\definecolor{darkred}{rgb}{0.5,0.0,0.0}
\definecolor{darkblue}{rgb}{0.0,0.0,0.9}
\definecolor{darkerblue}{rgb}{0.0,0.0,0.5}
\definecolor{darkgreen}{rgb}{0.0,0.5,0.0}
\definecolor{black}{rgb}{0.0,0.0,0.0}
\definecolor{brown}{rgb}{0.6,0.4,0.2}

\def\be{\begin{equation}}
\def\ee{\end{equation}}
\def\cL{\mathcal{L}}
\def\cO{\mathcal{O}}
\def\cD{\mathcal{D}}
\def\cM{\mathcal{M}}

\def\msbar{\overline{\text{MS}}}
\def\vcl{v_\text{cl}}

\newcommand{\mpl}{  M_{\text{Pl}} }

\def\K{K}
\def\X{X}
\def\Y{Y}
\def\muX{\mu_X}
\def\muY{\mu_Y}
\def\muZ{\mu_Z}
\def\muI{\mu_I}
\def\nn{\nonumber}
\def\e{\varepsilon}

\newcommand{\Vmin}{V_{\text{min}}}
\newcommand{\VLO}{V^{\text{LO}}}
\newcommand{\VNLO}{V^{\text{NLO}}}
\newcommand{\VNNLO}{V^{\text{NNLO}}}

\newcommand{\lhat}{\widehat{\lambda}}

\newcommand{\otwo}{ {\text{O}(2)} }

\newcommand{\fd}[2]{\parbox{#1}{\includegraphics[width=#1]{figs/#2}}}

\numberwithin{equation}{section}

\title{Consistent Use of Effective Potentials}

\author{Anders Andreassen\thanks{anders@physics.harvard.edu}  }
\author{William Frost\thanks{wfrost@physics.harvard.edu}  }
\author{Matthew D. Schwartz\thanks{schwartz@physics.harvard.edu} }

\affil{\emph{Department of Physics,
Harvard University, Cambridge, MA 02138, USA}}

\begin{document} 
\maketitle

\begin{abstract}
It is well known that effective potentials can be gauge-dependent while their values at extrema should be gauge-invariant.
Unfortunately, establishing this invariance in perturbation theory  is not straightforward, since contributions from arbitrarily high-order loops can be of the same size. We
show in massless scalar QED that an infinite class of loops can be summed (and must be summed) to give a gauge invariant value for the potential at its minimum.
In addition, we show that the exact potential depends on both the scale at which it is calculated and the normalization of the fields, but the vacuum energy does not.
Using these insights, we propose a method to extract some physical quantities from effective potentials which is self-consistent order-by-order in perturbation theory, including
improvement with the renormalization group.
\end{abstract}

\section{Introduction}
Effective actions provide a powerful organizing framework for quantum field theory. 
Often one computes an effective action by integrating out a massive particle, usually at fixed order in perturbation theory as in the 4-Fermi theory,
but sometimes to all orders in couplings but leading order in $\frac{1}{m^2}$, as in the Euler-Heisenberg Lagrangian~\cite{Heisenberg:1935qt, Schwinger:1951nm}.
In these cases, the effective action is an action just like the original one, but with a subset of its fields. Thus it can be used for the same calculations, such
as $S$-matrix elements, in the same way. An advantage of the effective action approach is that it is often easier to expand in $\frac{1}{m^2}$  and then calculate the
physical quantity than to perform the calculation before the expansion.

A more subtle use of effective actions occurs when one integrates out {\it all} the particles. Integrating out everything of course just produces a number (usually $\infty$). To get functional dependence one can instead integrate out all the particles in the presence of classical sources $J$ for each to get a functional $W[J]$ of those sources. The Legendre transform of $W[J]$ gives an effective action $\Gamma[\phi]= W[J] - J\phi$, which we call the 1PI effective action. In the 1PI effective action, the fields are classical background fields. One cannot perform loops using $\Gamma[\phi]$ since all the loops have already been performed in integrating out the particles. $\Gamma[\phi]$  nevertheless encodes all the classical and quantum physics of the original theory. This $\Gamma[\phi]$ is therefore an extremely powerful object. It is also extremely difficult to compute. 

These two types of effective action are of course closely related; for example, the 1PI effective action for QED reduces to the Euler-Heisenberg Lagrangian when restricted to a single electron loop and constant electromagnetic fields. It is nevertheless important to understand that the 1PI effective action does {\it not} function like the kind of integral over a local Lagrangian one usually uses in quantum field theory. $\Gamma[\phi]$ describes fluctuations in the presence of background sources, which are normally absent. The fields $\phi$ on which $\Gamma$ depends are classical fields, functions of space-time, not quantum operators acting on a Hilbert space. The value of the action at generic classical field values $\phi$ has no clear physical meaning. Indeed, its value depends on the way the calculation is performed: what gauge is chosen, the normalization of the fields, and even the scale at which the action is computed (see Section~\ref{sec:EA} below). On-shell quantities, such as $S$-matrix elements 
computed 
from the effective action do not depend on these conventions.
An important property of the effective action, or the effective potential to which it reduces for constant fields, is that its minimum gives the energy
of the true ground state of a theory. For this reason, effective actions provide unrivaled insight into spontaneous symmetry breaking~\cite{JonaLasinio:1964cw,Coleman:1973jx}.

Despite the unphysical nature of the effective action at generic field values, its gauge-dependence had elicited a surprising amount of consternation. Since the
gauge-dependence was first observed  forty years ago~\cite{Jackiw:1974cv}, papers are regularly written claiming that the effective potential should be defined in some other way, often through a field redefinition, so that it is manifestly gauge invariant.
The easiest way to make the potential gauge invariant is simply to pick a gauge: In the mid 1970's, Dolan and Jackiw argued that only unitary gauge ($\xi=\infty$) makes sense~\cite{Dolan:1974gu}
while S. Weinberg
insisted upon Landau gauge~\cite{Weinberg:1973ua} ($\xi=0$). 
Around the same time, Frere and Nicoletopoulos (based on observations of Fischler and Brout~\cite{Fischler:1974ue}) argued that one should use dressed fields which have no field strength renormalization~\cite{Frere:1974ia}, which is similar to Yennie gauge ($\xi=3$). 
In 1998, Tye and Vtorov-Karevsky argued that one should use an exact nonlinear field redefinition, replacing a scalar doublet  $\phi_1 + i \phi_2$ by a linear sigma model $\sigma e^{i \pi}$~\cite{Tye:1996au}. 
Others have tried using composite fields, such as $\phi^* \phi$~\cite{Buchmuller:1994vy}, working in a Hamiltonian formulation~\cite{Boyanovsky:1996dc}
or the Vikovisky-DeWitt~\cite{Lin:1998up} formalism, where the introduction of gauge-invariance (and therefore gauge non-invariance) is sidestepped completely. 
In 2014, Nielsen~\cite{Nielsen:2014spa} proposed a nonlinear, but perturbative, field redefinition determined by solving a differential
equation with a boundary condition at  $\xi=\xi_0$ to remove the dependence on $\xi$. 

The periodic frustration with the gauge-dependence of the effective potential seems to us somewhat misguided.
Gauge invariance of unphysical quantities is
commonplace. One is not normally bothered by the $\xi$-dependence of the photon two-point function or the anomalous dimension of a charged scalar field. Similarly, 
there is no reason to be dismayed about the gauge-dependence of the effective action.  
While field redefinitions can treat a symptom of the unphysical nature of the potential, they cannot make it any more physical. 
In fact, none of papers cited above seem to address the incongruity of both using the freedom in quantum field theory to perform field redefinitions and
claiming that a particular field redefinition is preferred.
Of course,
physical quantities, such as on-shell $S$-matrix elements must be gauge-invariant. Furthermore, gauge-invariance provides a very useful theoretical consistency check. Not only does it help verify that you did the relevant loop calculations correctly, but it can also verify that the quantity you are calculating could possibly be physical. Thus, for an object as unintuitive as the effective potential, gauge-dependence is in many ways actually desirable.

With these observations in mind, let us consider how gauge invariance of some physical quantities calculated from the effective potential have been
established in perturbation theory and beyond. Our focus in this paper is on scalar QED at zero-temperature. With a negative mass-squared for the scalar
this model is called the Abelian Higgs model. With zero mass we call it the Coleman-Weinberg model~\cite{Weinberg:1973ua}. 
General non-perturbative demonstrations that certain quantities calculated from the effective potential in these models should be gauge invariant
were provided by Nielsen~\cite{Nielsen:1975fs} and independently Fukuda and Kugo~\cite{Fukuda:1975di}. 
Quantities which
are expected to be gauge-invariant in these models  include the value of the potential at its extrema and the pole masses of the scalar and vector bosons
in the spontaneously broken phase.
We review some of the main results from these papers in Sections~\ref{sec:EA} and~\ref{sec:AHM}.

Although all of our new results pertain to the Coleman-Weinberg model, we devote Section~\ref{sec:AHM} to the Abelian Higgs model since it illustrates
a number of aspects of gauge-invariance not present when $m=0$. We try to provide a useful summary of this model since the literature is scattered and somewhat 
inconsistent. Also, issues which are not present in the Coleman-Weinberg model, such as unphysical gauge-dependent minimum of the effective potential and the presence of an additional dimensionful scale, provide a broader picture of some subtleties associated with gauge-invariance.

The main result of this paper concerns the extrema of the Coleman-Weinberg effective potential. We demonstrate that the minimum of the renormalization-group improved  potential in this model is gauge-invariant order-by-order in perturbation theory. To show this requires first an understanding of the appropriate perturbation
expansion.
Perturbation theory must respect the relation $\lambda \sim \hbar e^4$ which determines the minimum at 1-loop (see Section~\ref{sec:oneloopCW}). We
calculate all the 2-loop contributions in $R_\xi$ gauges proportional to $\hbar^2 e^6$ (Section~\ref{sec:twoloopCW}), as well as all the terms like $\hbar^{n+2} e^6\frac{e^{4n}}{\lambda^n}$  
coming from 3-loop and higher-order graphs (Section~\ref{sec:daisies}) which also contribute at order $\hbar^2 e^6$ when $\lambda \sim \hbar e^4$. Demonstrating gauge-invariance in perturbation theory also requires careful consideration of what dimensionful
scale the potential at the minimum should be expressed in terms of. The most obvious dimensionful scale $v=\langle \phi \rangle$ is a questionable candidate since
it is gauge-dependent itself. These issues are discussed in Section~\ref{sec:GIphys}, where we show explicitly that the instability scale $\Lambda_I$ defined so that $V(\Lambda_I)=0$ depends on $\xi$. 

Next we augment our perturbative calculation with a proposal for how perturbation theory in which gauge-invariance holds order-by-order can be used in a resummed effective potential. The challenge is that, first, resummation breaks the strict $\lambda \sim \hbar  e^4$ power-counting required for gauge-invariance and, second, that the renormalization-group kernel for the field strength renormalization, 
$\Gamma = \int \gamma\, d \ln \mu$ has gauge-dependence to all orders in any expansion. Our  solution exploits the invariance of physical quantities in the effective potential to the scale where the potential is calculated: we simply run the couplings towards the minimum before calculating the potential in perturbation theory. Then the field strength renormalization plays no direct role.

Although the potential at the minimum is gauge invariant, field values are gauge-dependent. For example, the expectation value of the field, $v=\langle \phi \rangle$, depends on $\xi$ (cf. Eq.~\eqref{v1AHM})
as does the {\it instabilitiy scale} $\Lambda_I$, defined as the field value where $V(\Lambda_I)=0$ (cf. Eq.~\eqref{LambdaIofxi}).
 Although the gauge-dependence of field values is often emphasized~\cite{Nielsen:1975fs,Loinaz:1997td,Gonderinger:2012rd,Patel:2011th} there are still a surprising number of papers which try to ascribe physical significance to values of $\phi$.
Sometimes the physical question one wants to answer is akin to when higher-dimension operators, for example from Planck scale physics, can significantly affect the potential. We present one gauge-invariant way to deal with these higher-dimension operators in Section~\ref{sec:HD}. 
Our conclusions
and implications of our work are discussed in Section~\ref{sec:conc}.

\section{Effective actions \label{sec:EA}}
We begin with a rapid introduction to effective actions and their gauge dependence.
In this section, $\phi$  generically denotes all the fields in a given theory (in the rest of the paper $\phi$ will denote a specific field in scalar QED).

A very useful object in quantum field theory is the 1PI effective action $\Gamma[\phi]$. When $\phi$ is independent of $x$, the effective action reduces to the effective potential
\be
\Gamma[\phi] = -\int d^4 x V(\phi) + {\text{derivative and nonlocal terms}}
\ee
The interactions in the effective action embody a sum over all of the off-shell 1PI graphs in the quantum theory. Since any graph can
be produced by sewing together 1P1 graphs, the effective action when used at tree-level can reproduce all the loops in the original theory.
Thus the (quantum) effective action, when used classically, reproduces the full quantum physics of a given classical theory. 
In compensation for this enormous power, one must sacrifice manifest locality: in general, $\Gamma[\phi] \ne \int d^4 x \cL[\phi(x)]$ for any $\cL$. Knowing
the effective action amounts to solving a theory. As this is generally impossible, one must resort to an approximation to the effective action and argue that this
approximation is valid for the computation of interest. 

Although one could compute the 1PI effective action by computing all 1PI graphs in a theory, this is generally  inefficient. Instead, one often computes the effective
action by relating it to the Legendre transform of the classical action and doing a background field expansion~\cite{Jackiw:1974cv}. We first define the functional
\be
W[J] \equiv -i \ln Z[J] = -i \ln \left[ \int \cD \phi e^{i S[\phi] + i\int d^4x \phi J}\right] \label{WJ}
\ee 
where $J(x)$ is an external current. The effective action is the Legendre transform of $W[J]$:
\be
\Gamma [ \phi ] = W[J_\phi] - \int d^4 x J_\phi \phi
\ee
Here $J_\phi$ denotes the value of the external current which would make a given $\phi$ solve the equations of motion. Equivalently, in the presence of a current $J_\phi$, the expectation value of the field is $\phi_J$:
\be
\phi_J = \langle J_\phi | \phi(x) | J_\phi \rangle = -i \frac{1}{Z} \frac{\partial Z[J] }{\partial J(x) }  \Big|_{J=J_{\phi_J}}=\frac{\partial W[J] }{\partial J(x) } \Big|_{J=J_{\phi_J}} \label{Jphi}
\ee
In this equation $J_\phi$ is the independent variable which determines $\phi_J$ as the expectation value of the field in its presence.
For example, when $J=0$, so that no current is turned on, then $\phi_0 = \langle \phi\rangle$ is the true vacuum in the theory. 
The conjugate relation is
\be
 \frac{\partial \Gamma[\phi]}{\partial \phi} \Big|_{\phi=\phi_J} = -J_\phi \label{dGammaJ}
\ee
Here $\phi_J$ is the independent variable to be chosen and used to compute $J_\phi$. Importantly, this equation implies that the true vacuum of the theory, $\phi_{0}$, is an extremum of the effective action.
There is an assumed one-to-one correspondence between $\phi_J$ and $J_\phi$, which is why the effective potential should always be convex.\footnote{The convexity issue is a thorny one~\cite{Weinberg:1987vp,Strumia:1998nf}, as the effective potential in the Standard Model appears to be perturbatively nonconvex. Conveniently,
 all the effective potentials we consider in this paper are  convex.}

The effective action is to be used classically, so that fields are always on-shell, satisfying their equations of motion. When $\phi$ is not an extremum of $\Gamma$, it represents
the expectation value of the field in the presence of a nonzero background current; it is the classical solution to a different system.
In other words, $\Gamma[\phi]$ encodes how the system responds to external currents for which Eq.~\eqref{Jphi} is satisfied.
That this Legendre transform produces the object whose vertices are 1PI graphs in the original theory is not obvious.
It is also not obvious that $\Gamma[\phi]$ can be computed by shifting the fields $\phi \to \phi + \phi_q$ and computing 1PI graphs involving $\phi_q$ with $\phi$ held fixed.
These three representations of $\Gamma[\phi]$ are explained in more detail in~\cite{Coleman:1973jx,Abbott:1981ke,Schwartz:2013pla}. 

\subsection{Gauge dependence \label{sec:GI}}
The main issue of concern in this paper is the gauge-dependence of the effective action.
The gauge-dependence arises because when the source $J$ is nonzero, the system has a nondynamical background charge density. Since
there is no corresponding coupling of this charge density to the photon, the Ward identity will consequently be violated~\cite{Jackiw:1974cv}.
In the special case when $J=0$, there is no charged background. Then $\phi_0=\langle 0 | \phi(x) |0\rangle$ is extremal and we have simply
\be
\Gamma[\phi_0] = W[0]= -i \ln Z[0] = -i \ln  \int \cD \phi e^{i S[\phi]} \label{GasZ}
\ee
This expression is gauge-invariant, since gauge-fixing modifies $Z[0]$ only by an infinite constant, which drops out of $\ln Z[0]$ (see~\cite{Fukuda:1975di} for a longer discussion with a careful treatment of the divergences in $Z[0]$).
Thus when $\frac{\partial \Gamma[\phi]}{\partial \phi}=0$, then the effective action is gauge invariant  by Eq.~\eqref{dGammaJ}. 

An alternative way to show gauge invariance at the extremum is by exploring how and when Ward identities are violated in the presence of a background current~\cite{Fukuda:1975di,Nielsen:1975fs}. In particular,
Ref.~\cite{Nielsen:1975fs} demonstrates that in Fermi gauges, where the gauge-fixing term is $-\frac{1}{2\xi} (\partial_\mu A_\mu)^2$, the effective potential satisfies
\be
\left[ \xi \frac{\partial  }{\partial \xi} 
+ C(\phi,\xi) \frac{\partial }{\partial \phi} \right]V(\phi,\xi) = 0
\label{Nielsen}
\ee
where $C(\phi,\xi)$ is some precisely well-defined object, independently calculable order-by-order in perturbation theory.
What this equation implies is that the gauge dependence of the effective potential can be compensated for by a rescaling of $\phi$. 
It also apparently implies that at an extremum, where $\frac{\partial  }{\partial \phi}V(\phi,\xi) = 0$,
the effective potential is automatically gauge invariant, $\frac{\partial  }{\partial \xi}V(\phi,\xi) = 0$, consistent with the simple argument in the previous paragraph. 

There are a number of unsettling features of the Nielsen identity. First of all, it is somewhat trivial, in the sense that one can always define a function $C(\phi,\xi)$ so that Eq.~\eqref{Nielsen} holds; one simply
defines $C(\phi,\xi)$ by Eq.~\eqref{Nielsen}. Indeed, this is a shortcut to calculating $C(\phi,\xi)$. This shortcut makes it clear that the Nielsen identity only guarantees that
 $\frac{\partial  }{\partial \xi}V(\phi,\xi) = 0$ at an extremum if $C(\phi,\xi)$ is finite at the extremum. However, as Nielsen himself observes~\cite{Nielsen:1975fs}, although $C(\phi,\xi)$ being infinite is ``unlikely for $\xi$ finite'' [p. 180], 
it is also true that ``$C(\phi,\xi)$ diverges in the one-loop approximation of the [Abelian] Higgs model'' [p. 182]. In fact, the potential at an extremum can be gauge-invariant with $C(\phi,\xi)$ finite or infinite. It
is also possible for $C(\phi,\xi)$ to be infinite at the extremum to all orders in perturbation theory but finite non-perturbatively.

For a concrete example, as we will see in Section~\ref{sec:GIphys}, in the Coleman-Weinberg model $C(\phi,\xi)$ is logarithmically divergent as $\phi$ approaches an extremum when the potential is expressed in terms of the expectation value $v=\langle \phi \rangle$. The origin of this divergence is that $v$ itself is gauge-dependent. If instead one expresses $V$ in terms of the $\msbar$ scale $\mu$, then $C(\phi,\xi)$ is finite at the minimum.
Thus, as we show in this paper, the value of the potential at the minimum is gauge-invariant order-by-order only in a carefully considered perturbation expansion. We do not find the Nielsen
identity, which is a non-perturbative statement, particularly helpful in directing this expansion.

\subsection{Rescaling and calculation scale invariance \label{sec:rescale}}

\begin{figure}
\begin{center}
	\includegraphics[width=0.48\textwidth]{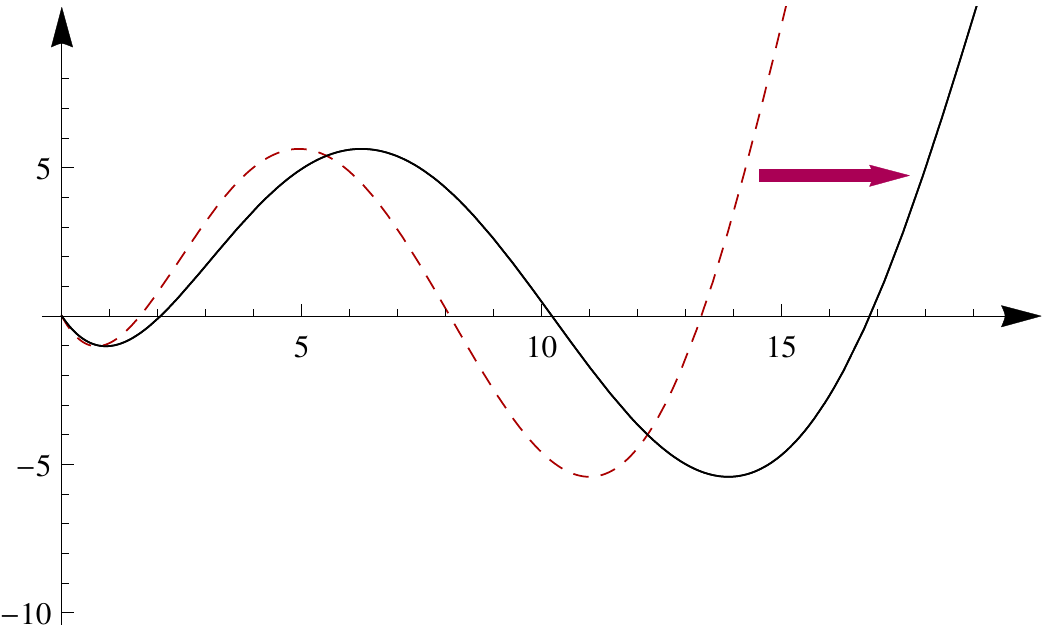}
\end{center}
\caption{Under the rescaling of the dependent variable, a function changes but its values at extrema do not \cite{Patel:2011th}.
This elementary mathematical fact explains why the effective potential can depend on the field normalization, but $\Vmin$ does not.
\label{fig:rescaling}
}
\end{figure}

Another feature of the potential at the minimum is that it is independent of field rescaling: $\phi \to \kappa\, \phi$ for any $\kappa$. This is not a deep observation, as the value of any function at a minimum is invariant under rescaling of the dependent variable (see Fig.~\ref{fig:rescaling}). More generally, the value of the potential at an extremum is independent of any field redefinition, $\phi \to \phi'(\phi)$. These are intuitive
features of quantum field theories, since the path integral is also invariant under field redefinitions, and the effective action in the true vacuum is defined as a path integral in Eq.~\eqref{GasZ}.
Indeed, gauge invariance of the action at extrema is a special case of general field redefinition invariance, since one can view gauge-transformations as field redefinitions. In doing so, however, one must allow 
for the possibility that if with one definition an extremum is at $A_\mu=0$, with another definition it may be with a nonzero and $x$-dependent expectation value for $A_\mu$.

A corollary of the above argument is that the potential away from its extrema {\it does} depend on how the field is normalized and defined. This is also obvious from Fig.~\ref{fig:rescaling}. Away from an extremum, the action describes the system in the presence of a current $J$. When one rescales the field $\phi$, the $ J \phi$ term with $J$ fixed breaks the invariance of the path integral under rescaling. Equivalently, from Eq.~\eqref{dGammaJ}, we see that $J_{\kappa\, \phi} = \frac{1}{\kappa} J_\phi$ so that when a field is rescaled, $\Gamma[\kappa \phi]$ gives the
least action in the presence of a rescaled current.

A number of authors have proposed that the gauge-dependence of the effective potential can be removed through a field redefinition~\cite{Fischler:1974ue,Frere:1974ia,Tye:1996au,Buchmuller:1994vy,Nielsen:2014spa}. For example, Tye and Vtorov-Karevsky argue that one should replace 
$\phi_1 + i \phi_2 \to \sigma \exp(i \pi)$~\cite{Tye:1996au}. Then $\sigma$ is a U$(1)$ singlet and so its source $J$ is neutral and the interaction $J\sigma$ in the
Lagrangian does not cause the Ward identity to
be violated. Although there is nothing wrong with this argument,
physical quantities, such as the value of the potential at its minimum, should be independent of field redefinitions. A field redefinition is in a sense similar to a gauge-choice. It does not make
the potential away from the minimum any more physical. Moreover, with this non-linear field redefinition, a renormalizable theory becomes nonrenormalizable and nominally straightforward calculations can  become drastically more complicated (try computing $\beta_\lambda$ at 1-loop in this theory). The point is that physics should be independent of field redefinitions, so one should choose a basis which makes calculations easiest, not one that makes unphysical quantities  more comforting.

An interesting and underappreciated fact about the effective potential  is that its form depends on the scale where it is computed. The effective potential satisfies an RGE:
\be
\Big(\mu \frac{\partial}{\partial \mu}
 - \gamma_i \phi_i \frac{\partial}{\partial \phi_i} 
 + \beta_i \frac{\partial}{\partial \lambda_i}\Big) V= 0
\label{genRGE}
\ee
This equation says that the explicit dependence on RG scale $\mu$ in the effective potential can be compensated by redefining the couplings $\lambda_i$, through their $\beta$ functions, and rescaling the fields $\phi_i$ based on their anomalous dimensions $\gamma_i$.
One can compute the potential to fixed order at a scale $\mu_0$, in terms of $\lambda_i(\mu_0)$ and $\mu_0$ and then evolve to some other scale $\mu$ by solving this equation simultaneously with the RGEs for the coupling constants. Call this method 1. Alternatively, one could evolve the couplings to $\mu$ and then compute the effective potential there. Call this method 2. 
The potentials at $\mu$ computed these two different ways will not agree: they differ because in method 1, the rescaling of $\phi_i$ from $\mu_0$ to $\mu$ is included, while in method 2, it is not. We show through an explicit example in Section~\ref{sec:CSI} that in the Coleman-Weinberg model, the two methods give potentials which indeed differ by $\ln\frac{\mu}{\mu_0}$ terms at next-to-leading order.

One can easily write a differential equation similar to Eq.~\eqref{Nielsen} for the scale-dependence:
\be
\left[ \mu_0 \frac{\partial  }{\partial \mu_0} 
-\gamma_i \phi_i \frac{\partial }{\partial \phi_i} \right]V(\phi,\mu_0,\mu) = 0
\label{ScInv}
\ee
where $\mu_0$ is the scale at which the potential is calculated before it is run to the scale $\mu$ by solving Eq.~\eqref{genRGE}. In a sense, this equation although trivial to derive is more useful than the Nielsen identity, 
since $\gamma_i$ cannot be infinite (unlike $C(\phi,\xi)$ in Eq.~\eqref{Nielsen}). It immediately indicates that at an extremum, the effective potential is $\mu_0$-independent. 
We use this calculation-scale invariance in Section~\ref{sec:RGimp} to show how a resummed effective potential can be computed so that it is gauge-invariant order-by-order in
a particular perturbation expansion.

For another perspective, recall that the vertices of the effective action encode (amputated) 1PI Green's functions. General amputated Green's functions satsify an RGE identical
to Eq.~\eqref{genRGE}, including the $\gamma$ term. Thus they also depend on the scale where they are calculated. In converting such Green's functions to $S$-matrix
elements, the LSZ reduction theorem instructs us to multiply by $\prod_i \sqrt{Z_i}$, where $\sqrt{Z_i}$ is the on-shell wave-function renormalization factor for external leg $i$. In $\msbar$,
these $Z$-factors are not 1 but are gauge and scale-dependent. Indeed, their $\mu$-dependence explicitly cancels the $\gamma$ term in Eq.~\eqref{genRGE} so that all scale dependence is
 $S$-matrix elements is compensated for by rescaling the couplings according to their $\beta$-functions. As the $\beta$-functions are gauge-invariant in $\msbar$, the $S$-matrix can then
be gauge-invariant at any scale.

In summary, the effective potential at general field values depends on an arbitrary calculation scale $\mu_0$. 
This is different from the usual dependence on the renormalization scale $\mu$
arising from a truncation to fixed order in perturbation theory. The effective potential has that dependence as well, but it depends on $\mu_0$ even non-perturbatively. The value of the potential at extrema is $\mu_0$-independent.
That the $\mu_0$-dependence drops out at extrema does not require field redefinitions or multiplication by the wavefunction renormalization factors included in going from Green's functions to $S$-matrix elements.

\section{Abelian Higgs Model \label{sec:AHM}}
This paper focuses on scalar QED, a renormalizable theory with two scalar fields $\phi_1$ and $\phi_2$, a photon, $A_\mu$ and an $\otwo$ symmetry. If the scalar doublet is given a negative mass-squared
in the classical potential, the model is called the Abelian Higgs model. This section is a review of results 
in~\cite{Dolan:1974gu,Aitchison:1983ns,DoNascimento:1987mn,Loinaz:1997td} with some details filled in and a few added comments. 
Establishing gauge-invariance in this model is simpler than in the Coleman-Weinberg model for the essential reason that there is spontaneous symmetry breaking already in the classical potential. Consequently, a traditional loop expansion is justified and the first non-trivial check occurs at 1-loop. Nevertheless, the Abelian Higgs model presents interesting features and subtleties complementary to those in the Coleman-Weinberg model,
which is why we discuss it here.

 We write the Lagrangian as
\be
\cL = -\frac{1}{4} F_{\mu\nu}^2 + \frac{1}{2}(\partial_\mu \phi_1 -e A_\mu \phi_2)^2 + \frac{1}{2}(\partial_\mu \phi_2 + e A_\mu \phi_1)^2 
 - V_0+ \cL_{\text{GF}} + \cL_{\text{ghost}} \label{sqedL}
\ee
with the classical potential
\be
V_0(\phi) = - \frac{1}{2}m^2\phi^2 + \frac{\lambda}{24}\phi^4
\ee
where 
\be
\phi^2 \equiv \phi_1^2+\phi_2^2
\ee
Since the theory has a global $\otwo$ symmetry rotating $\phi_1$ and $\phi_2$, the effective potential should only depend on $\phi$. 
When $m^2>0$, the classical potential is minimized when
\be
\vcl \equiv \langle \phi_1 \rangle_{\text{classical}} =\sqrt{\frac{6}{\lambda}} m 
\ee
We have used global $\otwo$ symmetry to put the vacuum expectation value (vev) of in $\phi_1$ with $\langle \phi_2 \rangle =0$.
Since we have normalized the vacuum energy so that $V_0(0)=0$, the value of the classical potential at the minimum is $V_0(\vcl) = -\frac{3}{2\lambda} m^4 = -\frac{\lambda}{24}\vcl^4$.

To compute quantum corrections to the vacuum energy, we need to fix a gauge. In gauge-fixing, as always, one looks for a deformation of the theory with
the fewest nasty features. To appreciate the relevant tradeoffs, it
is helpful to use a three-parameter family ($\xi$, $\Upsilon_1$ and $\Upsilon_2$) of gauges~\cite{Dolan:1974gu}. We write
\be
\cL_\text{GF} = -\frac{1}{2\xi} (\partial_\mu A^\mu + e \xi    \Upsilon_1 \phi_1+e \xi    \Upsilon_2 \phi_2  )^2
\ee
The associated Fadeev-Popov ghost Lagrangian is
\be
\cL_{\text{ghost}} = (\partial_\mu \bar{c})( \partial_\mu c)
 -  e^2 \xi \Upsilon_1 \phi_2 \,\bar{c} c
+ e^2 \xi \Upsilon_2 \phi_1\,\bar{c} c
\ee
where the ghosts $c$ and antighosts $\bar{c}$ are Grassmann-valued scalar fields.

For $\Upsilon_i=0$, these are the {\bf Fermi gauges}. Fermi gauges are appealing because they do not explicitly break the global $\otwo$ symmetry rotating $\phi_1$ and $\phi_2$
and because the ghosts decouple. A drawback of Fermi gauges is that there is uncanceled kinetic mixing between the gauge field and the scalars.
A more serious problem with Fermi gauges is that they can lead to singular intermediate results, as we will see, which make some calculations difficult.

The $R_\xi$ gauges~\cite{Rxi} are  $\Upsilon_1=\vcl$ and $\Upsilon_2=0$. 
Note that even in $R_\xi$ gauges, there is still kinetic mixing when  $\phi \ne \vcl$ and, except for Landau gauge, $\xi=0$, the
ghosts are not free. 
In the $R_\xi$ gauges, the global $\otwo$ symmetry is explicitly broken. 
This
breaking is not dangerous, since gauge-independent quantities must be independent of the gauge-fixing and should respect the $\otwo$ symmetry.
A more troubling fact about $\Upsilon_1=\vcl$ is that the {\it classical} minimum $\vcl$ is then hard-coded into the Lagrangian, when presumably only the true quantum minimum has any physical significance.
It would be even more troubling to allow $\Upsilon_i$ to be dynamical, for example with $\Upsilon_1 =\phi$; such gauges lead S. Weinberg to declare $R_\xi$ gauges problematic in all but Landau gauge, $\xi=0$~\cite{Weinberg:1973ua} (see~\cite{Fukuda:1975di} for a careful refutation of the argument in~\cite{Weinberg:1973ua}).
In the 3-parameter family we consider here, $\Upsilon_i$ are arbitrary, so $\vcl$ does not appear. 

Another feature of the Abelian Higgs model with this gauge fixing is that it has a spurious unphysical minimum. The gauge fixing induces a coupling between $\phi_1$ and $\phi_2$ in
the classical potential. Thus there is a solution for constant fields with $A_\mu = c = \bar{c} = 0$ and
\be
\phi_i = e\Upsilon_i \sqrt{\frac{6m^2}{\lambda e^2(\Upsilon_1^2 + \Upsilon_2^2)} + \frac{6 \xi}{\lambda} }
\ee
The potential on this solution has the value
\be
\Vmin=-\frac{3}{2\lambda}\left[m^2+e^2\xi (\Upsilon_1^2 + \Upsilon_2^2)\right]^2
\ee
which is gauge-dependent and unphysical. This solution, found by Dolan and Jackiw~\cite{Dolan:1974gu}, led them to argue for unitary gauge, $\xi=\infty$. As explained in~\cite{Fukuda:1975di} (see also~\cite{DoNascimento:1987mn}), the trouble
with this solution is not the gauge-dependence, but that, despite satisfying the Euler-Lagrange equations, it does not extremize the action. An extremal solution related to this one has a non-vanishing
and position dependent expectation value for $A_\mu$; the gauge-fixing term contributes at spatial infinity so the boundary terms cannot be dropped in deriving the Euler-Lagrange equations. In this sense
the 3-parameter gauge fixing provides what Ref.~\cite{Fukuda:1975di} calls a {\it bad gauge}. Note that there is nothing inconsistent about bad gauges, they are just rather inconvenient for doing calculations
because of the position-dependent extrema. If we stick to solutions close to the gauge-independent classical one, with $\phi_1 = \vcl$ and $\phi_2=0$, the position dependent solutions can be ignored. 

We can always use the $\otwo$ symmetry to keep $\langle \phi_2\rangle =0$ to all orders. If we do so, then we do not even need to turn on a background field for $\phi_2$, since it
will not contribute to the extremal solutions for $\phi_1$. To avoid the bad gauges discussed above, we should correspondingly take $\Upsilon_1=0$.
A more general $\otwo$-invariant condition is to only turn on the linear combination which has $\Upsilon_i \cdot \langle \phi_i \rangle =0$. This is essentially
what was done in~\cite{Aitchison:1983ns}. 
Then, the renormalized 1-loop effective potential in $\msbar$ with $m\ne 0$ is
\begin{multline}
V_1(\phi) = \frac{\hbar}{16\pi^2}  \left[
\frac{3}{4} M_A^4 \left( \ln\frac{M_A^2}{\mu^2} -\frac{5}{6}\right) 
 +\frac{1}{4} M_B^4 \left(\ln\frac{M_B^2}{\mu^2} -\frac{3}{2}\right) \right.\\
\left. 
-\frac{1}{2} M_G^4 \left( \frac{M_G^2}{\mu^2} - \frac{3}{2} \right)
+\frac{1}{4} M_{+}^4 \left(\ln\frac{M_{+}^2}{\mu^2} -\frac{3}{2}\right)
 +\frac{1}{4} M_{-}^4 \left(\ln\frac{M_{-}^2}{\mu^2} -\frac{3}{2}\right)
 \right] \label{V1m}
\end{multline}
where
\begin{align}
M_A^2 &= e^2 \phi^2 \\
M_B^2 &= \frac{\lambda}{2} \phi^2 - m^2 \\
M_G^2 &= \xi e^2( \Upsilon_1\phi_2 - \Upsilon_2 \phi_1)
\end{align}
are the contributions from $A_\mu$, $\phi$ and ghosts, respectively and
\begin{multline}
M_\pm^2 = \frac{\lambda}{12} (\phi^2 - \vcl^2)
+  \xi  e^2( \Upsilon_1\phi_2 - \Upsilon_2 \phi_1)\\
\pm \frac{1}{2} \sqrt{  \frac{\lambda}{6}\Big[\phi^2 - \vcl^2\Big] \Big[\frac{\lambda}{6}(\phi^2 - \vcl^2) - 4 \xi e^2 \phi^2 + 4 \xi e^2 ( \Upsilon_1\phi_2 - \Upsilon_2 \phi_1) \Big] } 
\end{multline}
come from the kinetic mixing. 

Then, assuming that the vev of $\phi_1$ gets a perturbative correction 
$v= \langle \phi_1 \rangle= \vcl +  v_1  +\cdots $ for some $v_1$ of order $\hbar$, we can evaluate
the potential at the minimum perturbatively:
\begin{align}
\Vmin &= V_0(v) +  V_1(v) + \cdots \\
&= V_0(\vcl) + v_1 V_0'(\vcl)+ V_1(\vcl) + \cdots\\
&= V_0(\vcl) + V_1(\vcl)  + \cdots+ \cO(\hbar^2)
\end{align}
where the omitted terms are $\cO(\hbar^2)$ and
 $V_0'(\vcl)=0$ has been used. 
The importance of doing an analytic expansion in $\hbar$ with an appropriate truncation before evaluating $\Vmin$ or other physical quantities was emphasized by Patel and Ramsey-Musolf in \cite{Patel:2011th} (see also \cite{Loinaz:1997td}).
Since $M_+^2 = M_-^2 = M_G^2= - \xi e^2 \Upsilon_2 \vcl$  when $\phi_1= \vcl$ and $\phi_2=0$, we find that $V(v)$ gauge-invariant to order $\hbar$. Explicitly,
\be
\Vmin = \vcl^4\left[-\frac{\lambda}{24} + \frac{\hbar}{16\pi^2}\left(-\frac{5 }{8} e^4 -\frac{1}{24}\lambda^2 + \frac{1}{36} \lambda^2 \ln \frac{\lambda \vcl^2 }{3 \mu^2} + \frac{3}{4} e^4 \ln \frac{e^2 \vcl^2}{\mu^2} \right)
 + \cO(\hbar^2)\right]
\ee
Thus the vacuum energy is manifestly gauge-independent. 

To determine $v_1$, we note that the minimum condition is
\be
0 = V_0'(v)+V_1'(v) +\cdots = V_0'(\vcl) + v_1 V_0''(\vcl) + V_1'(\vcl) + \cdots
\ee
using again that $V_0'(\vcl)=0$, we get
$v_1 = -\frac{V_1'(\vcl)}{V_0''(\vcl)}$ which gives
\begin{multline}
v_1= \vcl\frac{\hbar}{16\pi^2} \left[
 \frac{3 e^4}{\lambda} 
 + \frac{\lambda}{2} 
 - \frac{9 e^4}{\lambda}\ln\frac{e^2 \vcl^2}{\mu^2} 
 - \frac{\lambda}{2} \ln\frac{\vcl^2 \lambda}{3\mu^2}
   + \frac{e^2\xi}{2} \ln\frac{-e^2 \vcl \xi \Upsilon_2}{\mu^2} \right] 
   \\
  + \Upsilon_2 e^2 \xi\frac{\hbar}{16\pi^2}\left[-\frac{1}{2} +\ln \frac{-e^2 \xi \vcl \Upsilon_2}{\mu^2}\right]
\label{v1AHM}
\end{multline}
This expression manifestly depends on both $\xi$ and $\Upsilon_2$. Dangers of assigning physical significance to $\langle \phi \rangle$ have been emphasized in~\cite{Loinaz:1997td,Patel:2011th} and elsewhere.

Note that $v_1$ is singular in Fermi gauges, $\Upsilon_i \to 0$, but not in Landau gauge $\xi\to 0$. The singularity is an infrared divergence associated with the ghosts becoming massless. As pointed out in~\cite{Nielsen:1975fs} and \cite{Aitchison:1983ns}, the function $C(\phi,\xi)$ in the Nielsen identity also has an infrared divergence in Fermi gauge. As discussed in Section~\ref{sec:GI}, that $C(\phi,\xi)$ can be infinite calls into question the usefulness of the Nielsen identity -- as we have seen, the potential at the minimum is gauge-invariant for any choice of $\Upsilon_2$ and  $\xi$. 

At this point, we can conclude that everything works swimmingly: the vacuum energy is gauge invariant but the vev is not, exactly as foretold by the Nielsen identity.
At the risk of being too cautious, it is perhaps worth adding that in the Abelian Higgs model, $\Vmin$ is not actually observable -- the true renormalized vacuum energy has an arbitrary subtraction associated with the cosmological constant. Even if the vacuum energy can be measured (for example, by weakly coupling this model to gravity), one still needs a scheme to measure the difference $\Delta V=\Vmin-V(0)$ between this energy and the energy at $\phi=0$. This difference, as a function of $\vcl$, $e$ and $\lambda$ is what we have calculated. Even then, 
although $\Delta V$ is independent of gauge, it is expressed in terms of $\vcl = \sqrt{\frac{6}{\lambda}} m$ which is itself unphysical. Indeed, even $m$ is not physical: it is not the mass of the scalar in the vacuum. That mass, which we denote $m_S$, itself gets corrections of 
order $\hbar$ from self-energy graphs. The gauge-dependent part of these corrections at order $e^2 \lambda$ have been calculated in~\cite{Aitchison:1983ns}, with the result that $\frac{d}{d\xi} m_S =0$.
Therefore $\Delta V/m_S^4$ is gauge independent and only a function of the $\msbar$ couplings $e$ and $\lambda$. This ratio is in principle observable, if $\Delta V$ is, and depends only on other observables, $e$ and $\lambda$. With this long-winded disclaimer, we can now conclude that quantum corrections in effective potential can make (in principle) testable experimental predictions.

In summary, in the Abelian Higgs model both $\Delta V=\Vmin-V(0)$ and $m_S$  are separately independent of the gauge parameters $\xi$ and $\Upsilon_2$
 at order $\hbar$ when expressed in terms of the classical expectation value of $\phi$, $\vcl$. That $\Delta V$ and $m_S$ are separately gauge invariant  is not strictly necessary. All that is required is that the gauge-dependence cancels in the ratio $\Delta V/m_S^4$ which is a calculable function of the observables $e$ and $\lambda$.  As we will see, in the Coleman-Weinberg model, where the classical theory is scaleless, such considerations are critical.
We also found that  the expectation value of the field $v= \langle \phi \rangle$ is {\it not} gauge invariant at order $\hbar$. In the Fermi gauges, $\Upsilon_i=0$, $v$ even has an infrared divergence.

\section{Coleman-Weinberg Model \label{sec:oneloopCW}}
Now let us turn to the main subject of this paper, scalar QED with a massless scalar, also known as the Coleman-Weinberg model. 
The Lagrangian is as in Eq.~\eqref{sqedL}, with
\be
V_0=  \frac{\lambda}{24}\phi^4 \label{V0}
\ee
and the gauge-fixing term is
\be
\cL_\text{GF} = -\frac{1}{2\xi} (\partial_\mu A^\mu)^2
\ee
These can be thought of as  $R_\xi$ gauges (since $\vcl=0$ in this theory) or Fermi gauges. They correspond to $\Upsilon_1=\Upsilon_2=0$ in the 3-parameter gauge family discussed above, and so the ghosts decouple and can be ignored. 
There is unavoidable kinetic mixing between $\phi$ and $\partial_\mu A_\mu$ in this theory, but as in the Abelian Higgs model, this is an inconvenient but not insurmountable complication. 

The renormalized 1-loop effective potential in $\msbar$ in this
theory is the $m\to 0$ limit of Eq. \eqref{V1m}:
\begin{multline}
V_1(\phi) = \phi^4\frac{\hbar}{16\pi^2}  \left[
\frac{3}{4} e^4  \left( \ln\frac{e^2 \phi^2}{\mu^2} -\frac{5}{6}\right) 
 +\frac{\lambda^2}{16}  \left(\ln\frac{\lambda\phi^2}{2\mu^2} -\frac{3}{2}\right)
  \right.\\
\left. 
+\left(\frac{\lambda^2}{144}-\frac{1}{12} e^2 \lambda  \xi\right)\left(\ln\frac{\phi^2}{\mu^2} - \frac{3}{2} \right)
+\frac{1}{4} \K_{+}^4 \ln \K_{+}^2
+\frac{1}{4} \K_{-}^4 \ln \K_{-}^2
 \right] \label{V1}
\end{multline}
with
\be
\K_\pm^2 = \frac{1}{12}\left(\lambda  
\pm  \sqrt{  \lambda^2 -24\lambda e^2 \xi   } \right)
\label{kpmdef}
\ee
The relation 
\be
\K_+^4 + \K_-^4 = \frac{\lambda^2}{36}-\frac{1}{3} e^2 \lambda  \xi
\ee
has been used to simplify the 1-loop potential.

The tree-level potential in this model has only a single minimum, at $\phi=0$, where the $\otwo$ symmetry is unbroken. For there to be a minimum at 1-loop,
the corrections must be large enough to turn over the potential. For $\lambda$ small, so that the theory is perturbative, this is only possible if the 
$\frac{\hbar}{16\pi^2} e^4$ term is as large as the tree-level $\frac{\lambda}{24}$ piece. So let us assume $\lambda \sim \frac{\hbar}{16\pi^2} e^4$
and that there is a minimum at some scale $v$. Then the condition for the minimum, $V'(v) = 0$ provides a precise relationship between $\lambda$ and $e$:
\be
\lambda = \frac{\hbar}{16\pi^2} e^4\left(6-36 \ln\frac{e v}{\mu} \right) + \cO(e^6) \label{lambdacond}
\ee

As discussed in~\cite{Coleman:1973jx}, to understand this equation one must appreciate dimensional transmutation.
In scalar QED, the only scale is the scale $\mu$ at which the couplings are defined. $\mu$ is arbitrary, so we may as well take $\mu=v$. Then Eq.~\eqref{lambdacond} reduces to
\be
\lambda = \frac{\hbar}{16\pi^2} e^4\left(6-36 \ln e \right) + \cO(e^6) \label{lambdacond2}
\ee
This equation should be thought of as a condition on $\lambda=\lambda(\mu)$ and $e=e(\mu)$: the minimum occurs at the scale $v=\mu$ where Eq.~\eqref{lambdacond2} holds.
Of course, $v$ will get corrections and, as we will see, is gauge-dependent (unlike $\mu$). But at least at 1-loop, this is an acceptable  way to think about the
minimum in the effective potential in scalar QED.

Since $\lambda$ and $e$ can be anything, it is natural to wonder whether Eq.~\eqref{lambdacond2} requires some kind of finite tuning. As explained in~\cite{Coleman:1973jx}
it does not. The evolution of $e$ and $\lambda$ are determined by the $\beta$ functions:
\be
\mu \frac{d}{d \mu} e = \beta_e,\quad 
\mu \frac{d}{d \mu} \lambda = \beta_\lambda,\quad 
\ee
where, at 1-loop,
\begin{align}
\beta_e & = 
\frac{\hbar}{16\pi^2}\left( \frac{e^3}{3} \right) + \cdots \\
 \beta_\lambda & =
\frac{\hbar}{16\pi^2}\left( 36 e^4-12 e^2 \lambda +\frac{10 \lambda ^2}{3}\right)
\end{align}
The key feature of these equations is that $e$ is multiplicatively renormalized ($e=0$ is a fixed point of the RG flow) while $\lambda$ can get an additive
correction even at $\lambda=0$. What this means is that if $\lambda$ and $e$ start off small, $e$ runs logarithmically, but $\lambda$ will grow at ever increasing
rate until it hits a Landau pole. Indeed, the exact solutions to the 1-loop RGEs are~\cite{Coleman:1973jx}
\be
e^2(\mu) = \frac{e^2(\mu_0)} {1- \frac{e^2(\mu_0)}{24\pi^2}\ln\frac{\mu}{\mu_0} } \label{e2mu}
\ee
which expresses $e(\mu)$ in terms of $e$ at some reference scale $\mu_0$ and
\be
\lambda(\mu) = \frac{ e^2(\mu)}{10} \left[19+  \sqrt{719}\tan \left(\frac{\sqrt{719}}{2} \ln\frac{ e(\mu)^2}{C} \right)\right] \label{lambdamu}
\ee
with $C$ an integration constant which can be traded for $\lambda(\mu_0)$. The tangent in  $\lambda(\mu)$ implies that as $e^2$ changes
by a factor of $\approx 1.2$, $\lambda$ will go from $-\infty$ to $\infty$. In particular, if $\lambda$ and $e$ are small, there will {\it always}
be a point where Eq.~\eqref{lambdacond2} is satisfied.

At the minimum, we find
\be
\Vmin = v^4 \frac{\hbar}{16\pi^2} e^4\left(-\frac{3}{8} \right)+ \cO(e^6)
\label{Vmin1}
\ee
This is gauge-invariant, simply because the $e^4$ terms in $V_1(\phi)$ are gauge-invariant.

The first non-trivial check on the gauge invariance of $V(v)$
requires the terms in the effective potential of order $\hbar^2 e^6$, with $\lambda$ counting as order $\hbar e^4$, and $\ln e$ and $\ln \lambda$
counting as order $e^0$. In scalar QED, each loop comes with a factor of $\hbar e^2$ or $\hbar \lambda$ from the vertices, so
$\hbar e^2 \lambda$ terms come from 1-loop graphs and $\hbar^2 e^6$ terms from 2-loop graphs. Thus we need at least the 2-loop Coleman-Weinberg potential.
In addition, effective potential calculations involve an infinite number of background field insertions, which can be conveniently resummed into dressed
propagators with $e$- and $\lambda$-dependent masses. These propagators allow for graphs to have extra factors of $e^2/\lambda$ in them.
For example, a 3-loop graph gives a term proportional to $\hbar^3 e^{10}/\lambda$ which
also scales like $\hbar^2  e^6$ when $\lambda \sim \hbar e^4$. These higher-loop terms are discussed in Section~\ref{sec:daisies}.

\section{The two-loop Coleman-Weinberg potential \label{sec:twoloopCW}}
The complete 2-loop potential in scalar QED with gauge-dependence does not appear in the literature, as far as we are aware. The
2-loop potential in Landau gauge is known in the complete Standard Model~\cite{Ford:1992pn} and for general renormalizable theories~\cite{Martin:2001vx}. Kang computed the terms in the 2-loop potential at
order $e^6$
that depend on $\ln\frac{\phi^2}{\mu^2}$, but with a subtraction scheme slightly different from $\msbar$ \cite{Kang:1974yj}.
For our analysis, we require also the $\mu$-independent part of the effective potential at order $e^6$.
We will also need an infinite series of other graphs, the {\it daisy} graphs, which are discussed
in Section~\ref{sec:daisies}, but let us start with the 2-loop computation.
To compute the effective potential at this order, there is a counterterm contribution and a contribution from 2-loop graphs. 
We work in $d=4-2\e$ dimensions throughout, and the scale $\mu$ is the $\msbar$ renormalization scale with implicitly contains the $\ln 4\pi$ and $\gamma_E$ factors. 

\subsection{Counterterm contribution}

The counterterm contribution can be extracted from the $\cO(\e)$ part of the 1-loop potential multiplying the $\frac{1}{\e}$ poles
in the renormalization factors. In $d=4-2\e$ dimensions, the 1-loop potential can be written as
\be
V_1 = \phi^4\Big[ (3-2\e) f(e^2)+f(\frac{\lambda}{2}) + f(\K_+^2) +f(\K_-^2)\Big] + \cO(\e^2)
\ee
where
\be
f(x) = -\frac{x^2}{64\pi^2} \left[ \frac{1}{\e} + \frac{3}{2} - \ln \frac{x \phi^2}{\mu^2}
+ \frac{\e}{2}  \left( \ln\frac{x\phi^2}{\mu^2} - \frac{3}{2} \right)^2+\e\left(\frac{5}{8} + \frac{\pi^2}{12} \right) 
\right]
\ee
with $\K_\pm$ defined in Eq.~\eqref{kpmdef}. The 1-loop renormalization factors are
\begin{align}
Z_\phi &= 1+ \frac{\hbar}{16\pi^2} \frac{1}{\e}(3-\xi)e^2 + \cdots, \nn \\
Z_e &= 1+ \frac{\hbar}{16\pi^2} \frac{1}{\e}\frac{e^2}{6} +\cdots\nn \\
\lambda Z_\lambda &= \lambda + \frac{\hbar}{16\pi^2} \frac{1}{\e} \left( 18 e^4  - 6 e^2\lambda +\frac{5}{3}\lambda^2 \right) + \cdots
\end{align}
Replacing $e \to Z_e e$, $\lambda \to Z_\lambda \lambda$ and $\phi \to \sqrt{ Z_\phi} \phi$ and expanding to order $\hbar^2 \e^0$ gives the counterterm contribution to the 2-loop potential:
\begin{multline}
V_2^{\text{ct}} = \left(\frac{\hbar}{16\pi^2}\right)^2 \phi^4 e^6\left[(-10+6 \xi) \ln^2\frac{e\phi}{\mu}
+ \left(\frac{10}{3} - 4 \xi + \frac{3}{2} \xi \ln \frac{\lambda \xi}{6 e^2} \right) \ln\frac{e\phi}{\mu}
\right.\\
\left.
-\frac{5}{3} - \frac{5 \pi^2}{12} + 2\xi + \frac{\pi^2}{16}\xi - \frac{3}{4} \xi \ln \frac{\lambda \xi}{6 e^2}+\frac{3}{16}\ln^2 \frac{\lambda \xi}{6 e^2} 
\right]  + \cO(e^4 \lambda,e^2\lambda^2,\lambda^3) \label{V2ct}
\end{multline}

\subsection{2-loop contribution}
The 2-loop contribution to the next-to-next-to-leading order effective potential can be computed from Feynman diagrams with dressed propagators (called {\it prototype diagrams} in~\cite{Coleman:1973jx}). 
The most straightforward way to compute the effective potential is to turn on a background field for $\phi_1$ by replacing $\phi_1 \to \phi_1 + \phi$ in the classical Lagrangian.
After substituting this into the classical Lagrangian and dropping the terms linear in propagating
fields $\phi_1$, $\phi_2$ and $A_\mu$, we compute the propagators by diagonalizing the kinetic terms, treating the background field $\phi$ as constant. 
The resulting propagators are:
\begin{align}
1~\fd{2cm}{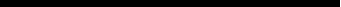}~1 ~~=~~ D_{11}(k) &= \frac{i}{k^2 - \frac{\lambda}{2} \phi^2} \label{D11def}\\
2~\fd{2cm}{scalarline}~2 ~~=~~ D_{22}(k) &= i \frac{k^2 - e^2 \xi \phi^2}{(k^2-\phi^2\K_+)(k^2-\phi^2\K_-)}\\
\mu~\fd{2cm}{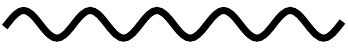}~\nu ~~=~~ \Delta_{\mu\nu}(k) &= -i \left[ \frac{1}{k^2 -e^2 \phi^2}\left(g_{\mu\nu} - \frac{k_\mu k_\nu}{k^2} \right) + \frac{\xi (k^2 - \frac{1}{6} \lambda\phi^2)}{(k^2-\phi^2\K_+)(k^2-\phi^2\K_-)} \frac{k_\mu k_\nu}{k^2}\right]
\end{align}
with $\K_\pm$ given in Eq.~\eqref{kpmdef} and
\be
\mu~\fd{2cm}{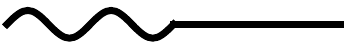}~2 ~~=~~ T_{\mu2} (k) = \frac{\xi e \phi}{(k^2-\phi^2\K_+)(k^2-\phi^2\K_-)} k_\mu \label{Tmu2def} 
\ee
This last propagator comes from the kinetic mixing between $\phi_2$ and $A_\mu$ in the presence of a background field for $\phi_1$.  The asymmetry between $D_{11}$ and $D_{22}$ arises because we are only turning on a background for $\phi_1$ for simplicity. The full effective potential of course also depends on background values for $\phi_2$. However, since the effective potential has a global $\otwo$ symmetry, it is enough to turn on a $\phi_1$ background for the questions we address in this paper.

The propagators in Eqs.~\eqref{D11def} to \eqref{Tmu2def} are
effective propagators, where all of the flipping between scalars and longitudinal vector bosons that can occur between interactions has already been resummed. In this way, an infinite set of diagrams is replaced by a single one:
\be
\fd{1.5cm}{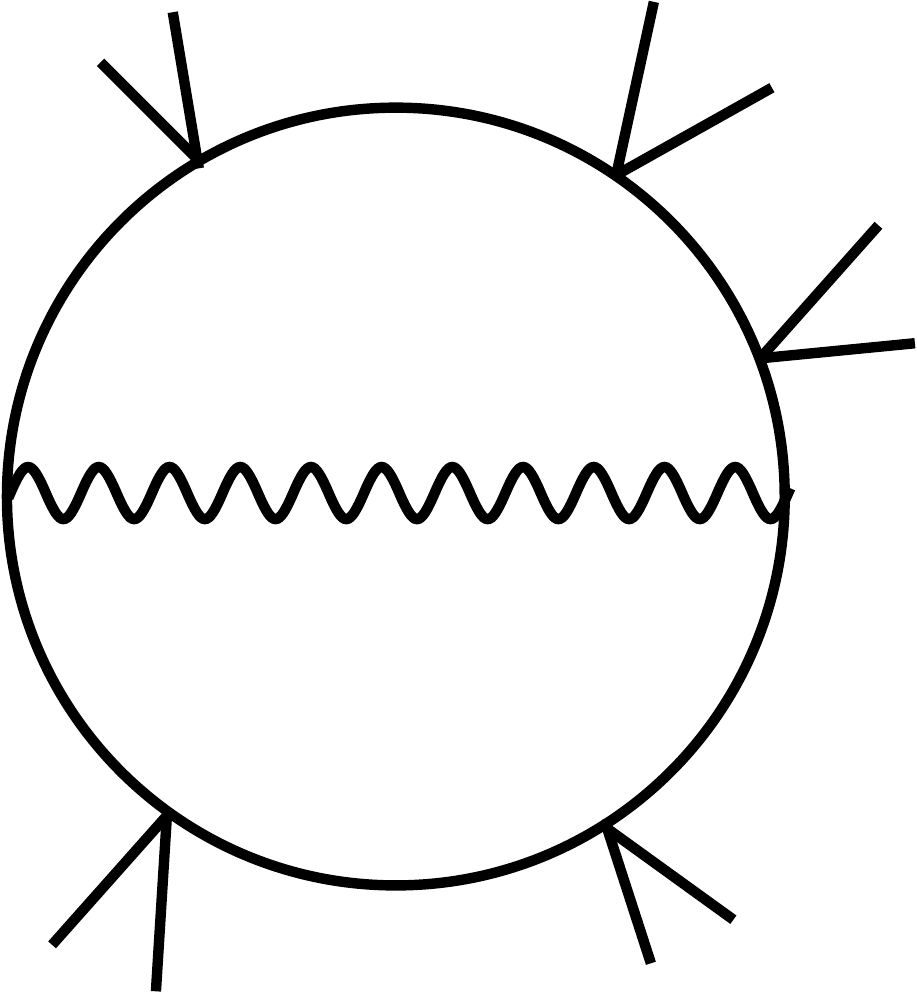} ~~+~~ \fd{1.5cm}{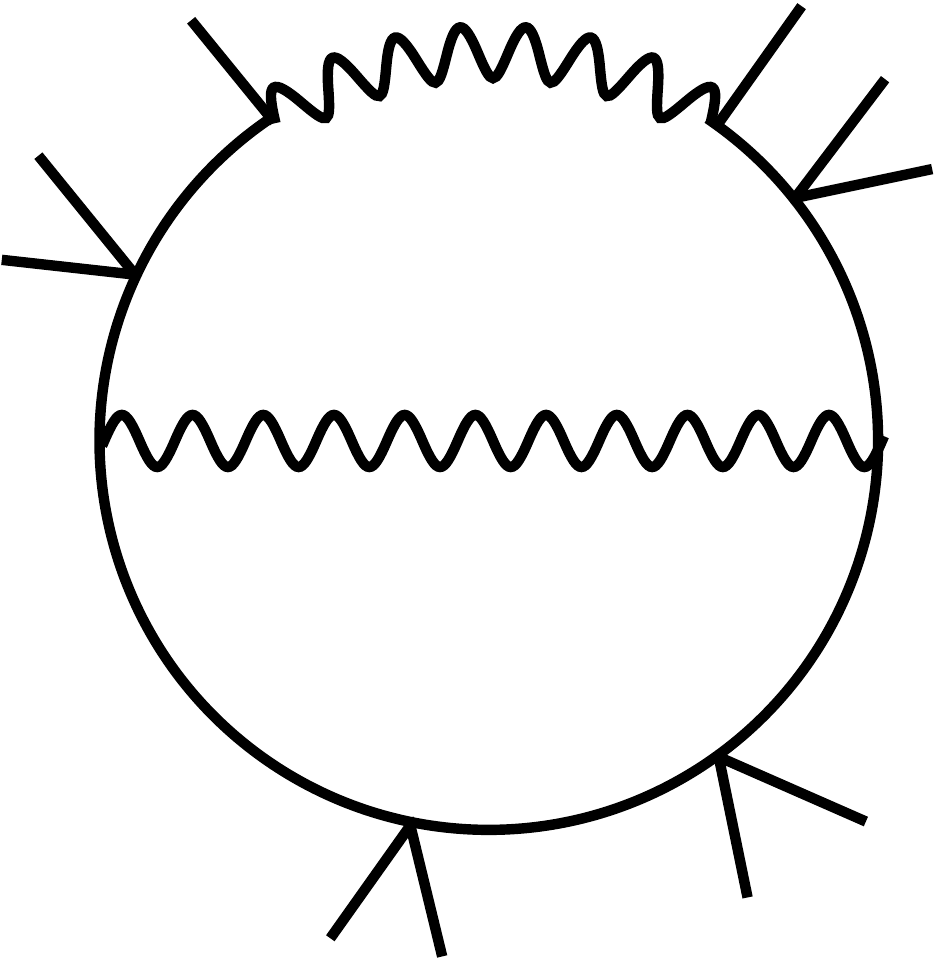}  ~~+~~ \fd{1.5cm}{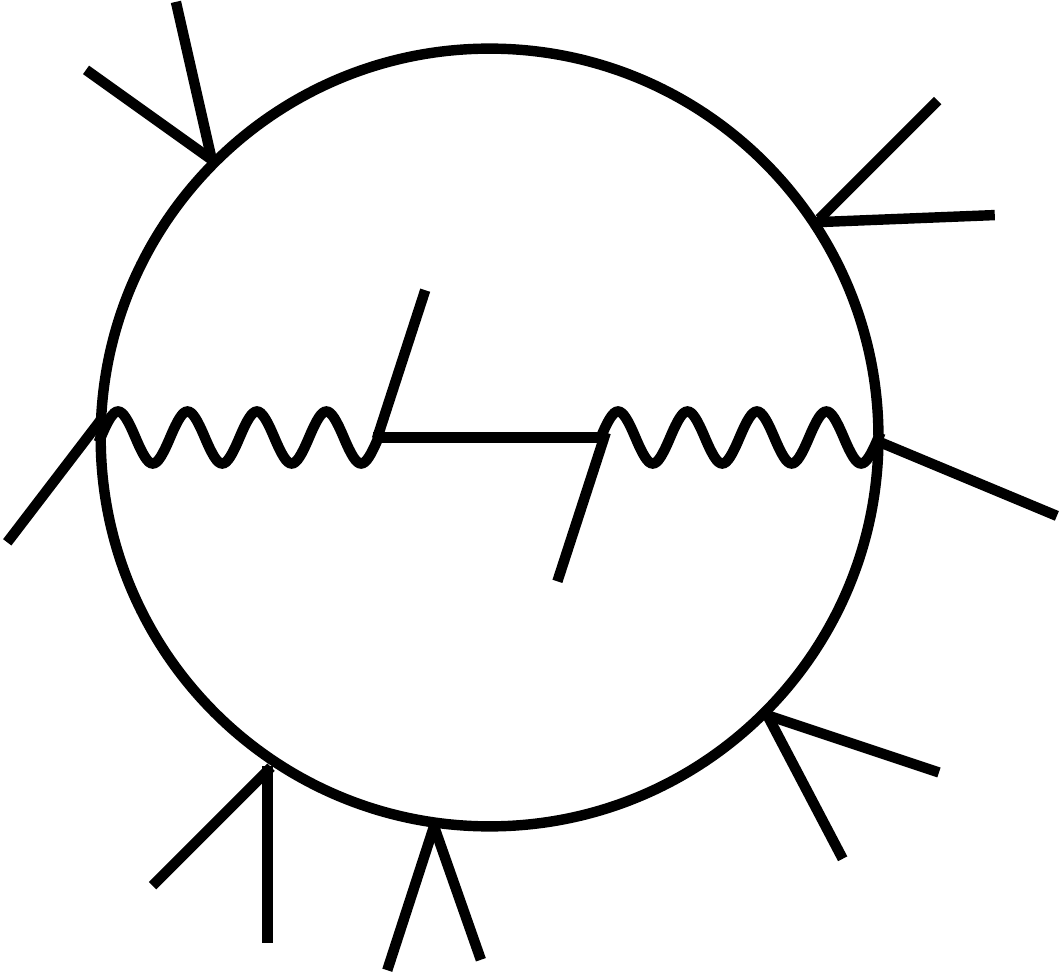} ~~+~~ \fd{1.5cm}{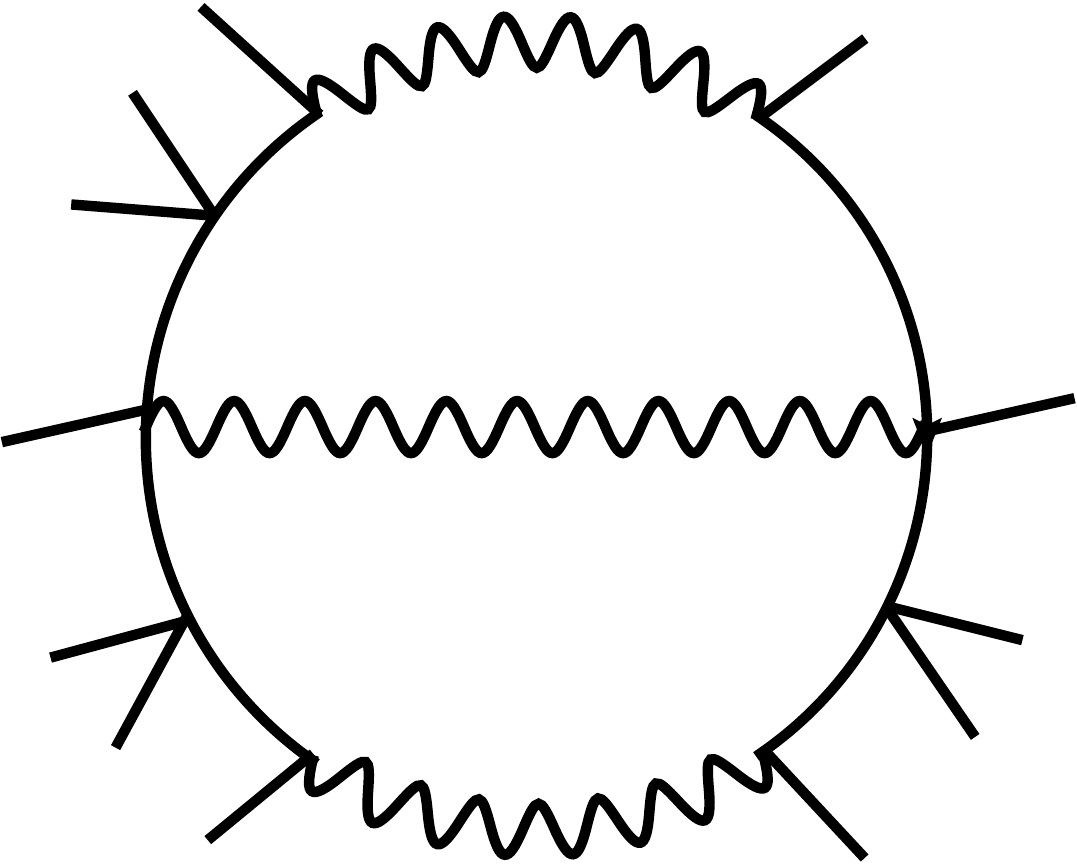} ~~+~~ \cdots ~~ \longrightarrow ~~ \fd{1.2cm}{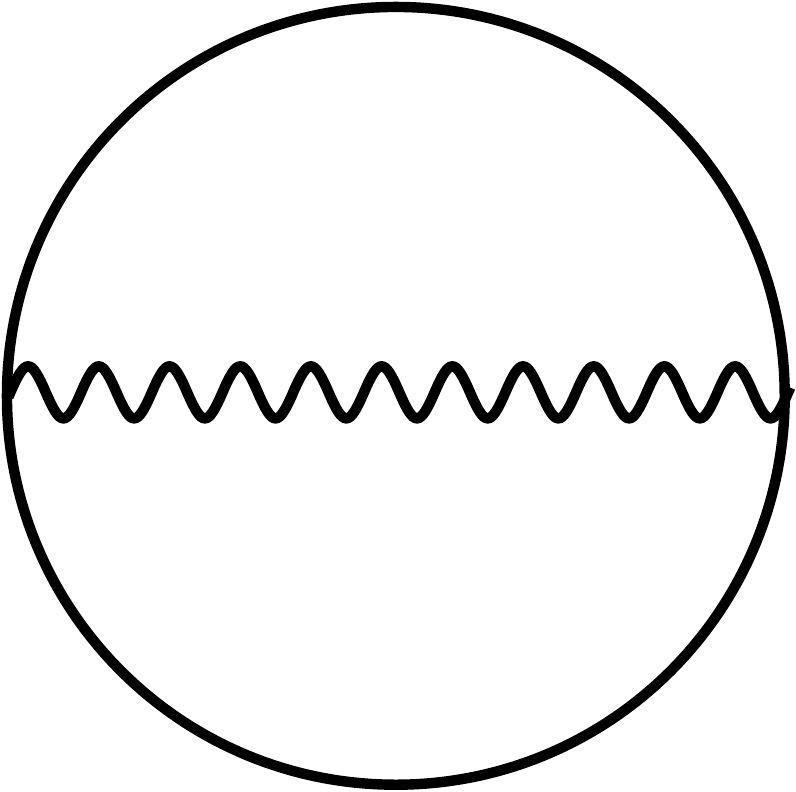}
\ee
Diagrams where there is one flip between vertices are still allowed (for example, see Fig.~\ref{fig:twoloops} below). 
The hair on the left-hand-side diagrams indicates background field insertions which are implicit in prototype diagrams,
even without kinetic mixing.

There are five types of interactions:
\be
 \hspace{-2mm}
\resizebox{10mm}{!}{
\parbox{10mm}{
\begin{tikzpicture}[]
\node (label) at (0,0){ \fd{1.5cm}{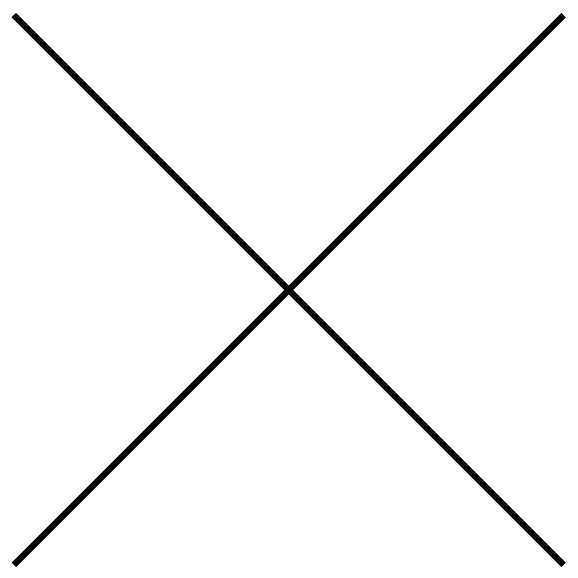} };
\node[rotate=0,anchor=east] at (-0.65,0.7) {$i$};
\node[rotate=0,anchor=east] at (-0.65,-0.5) {$j$};
\node[rotate=0,anchor=east] at (1.15,-0.5) {$k$};
\node[rotate=0,anchor=east] at (1.15, 0.7) {$l$};
\end{tikzpicture}
}}
\hspace{9mm}
 = -i\frac{\lambda}{3}(\delta_{ij}\delta_{kl}+\delta_{ik}\delta_{jl}+\delta_{il}\delta_{jk})\
 \hspace{8mm}
\resizebox{10mm}{!}{
\parbox{10mm}{
\begin{tikzpicture}[]
\node (label) at (0,0){ \fd{1.5cm}{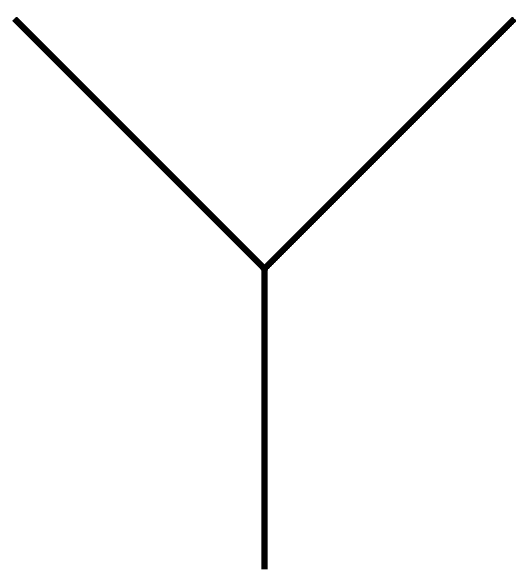} };
\node[rotate=0,anchor=east] at (-0.65,0.7) {$i$};
\node[rotate=0,anchor=east] at (1.15, 0.7) {$j$};
\node[rotate=0,anchor=east] at (0,-0.5) {$k$};
\end{tikzpicture}
}}
\hspace{9mm}
= - i \frac{\lambda}{3}\phi(\delta_{ij}\delta_{k1}+\delta_{jk}\delta_{i1}+\delta_{ki}\delta_{j1})
\ee
\be
 \hspace{-2mm}
\resizebox{10mm}{!}{
\parbox{10mm}{
\begin{tikzpicture}[]
\node (label) at (0,0){ \fd{1.5cm}{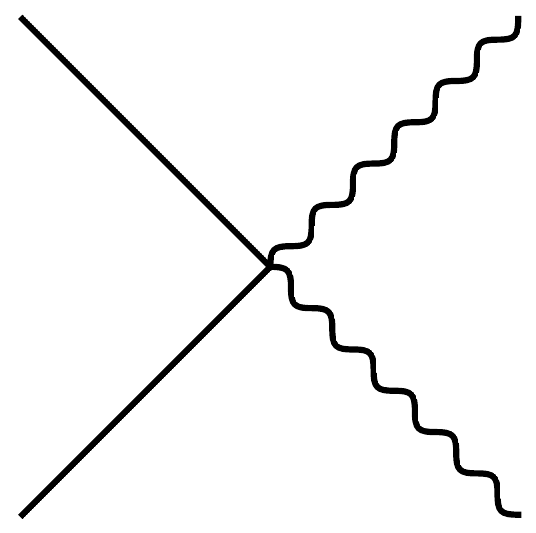} };
\node[rotate=0,anchor=east] at (-0.65,0.7) {$i$};
\node[rotate=0,anchor=east] at (-0.65,-0.5) {$j$};
\node[rotate=0,anchor=east] at (1.15,-0.5) {$\nu$};
\node[rotate=0,anchor=east] at (1.15, 0.7) {$\mu$};
\end{tikzpicture}
}}
 \hspace{8mm}
= 2i e^2 \delta_{ij}g_{\mu\nu}
 \hspace{8mm}
\resizebox{10mm}{!}{
\parbox{10mm}{
\begin{tikzpicture}[]
\node (label) at (0,0){ \fd{1.5cm}{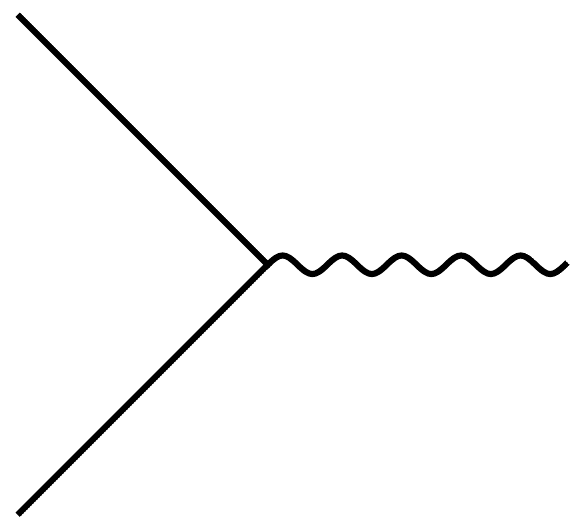} };
\node[rotate=0,anchor=east] at (-0.65,0.7) {$i$};
\node[rotate=0,anchor=east] at (-0.65,-0.5) {$j$};
\node[rotate=0,anchor=east] at (0.9,0.3) {$\mu$};
\end{tikzpicture}
}}
\hspace{8mm}
= e\epsilon_{ij}(k_i^\mu+k_j^\mu)
 \hspace{8mm}
\resizebox{10mm}{!}{
\parbox{10mm}{
\begin{tikzpicture}[]
\node (label) at (0,0){ \fd{1.5cm}{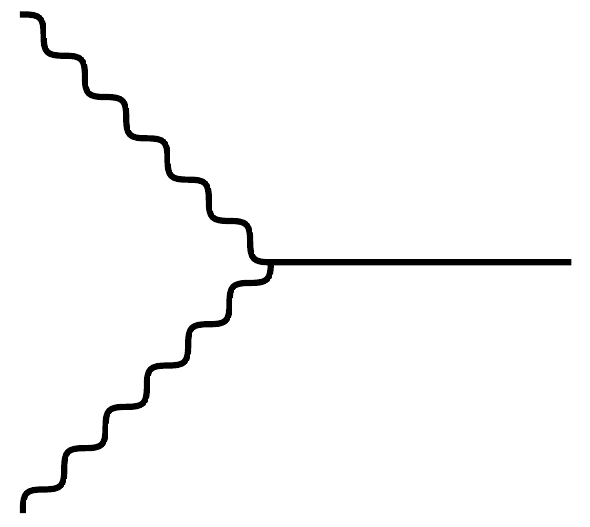} };
\node[rotate=0,anchor=east] at (-0.65,0.7) {$\mu$};
\node[rotate=0,anchor=east] at (-0.65,-0.5) {$\nu$};
\node[rotate=0,anchor=east] at (0.9,0.3) {$j$};
\end{tikzpicture}
}}
\hspace{10mm}
= 2ie^2\phi \delta_{j1}g_{\mu\nu}
\ee
where the solid lines can be either $\phi_1$ or $\phi_2$. 
These Feynman rules can also be found in~\cite{Kang:1974yj} after rescaling $\lambda \to \frac{\lambda}{3}$ to match our normalization,
with an extended discussion of their derivation in Appendix A of that reference.
Diagrams constructed with these Feynman rules were dubbed {\it prototype diagrams} in \cite{Coleman:1973jx} to emphasize that they represent
an infinite number of diagrams with arbitrary numbers of background fields insertions.

The effective potential is computed with these Feynman rules by evaluating vacuum energy contributions: diagrams with no external propagating fields. Each such diagram corresponds to an infinite number of background field insertions. One can also compute terms in the effective action which have derivatives, such as corrections to the kinetic term $(\partial_\mu \phi)(\partial_\mu \phi)$. To do so, one inserts finite momentum $p^\mu$ into a single background field line. The procedure is described in~\cite{Fraser:1984zb}. The 1-loop effective potential is a special case: the vacuum diagrams are quadratically divergent. A trick to computing it is to add external legs with $p^\mu =0$. 
Inserting one such leg produces the first derivative of the potential, $V'(\phi)$. With two legs gives the second derivative, $V''(\phi)$ and so in. Conversely, the second derivative of the potential $V''(\phi)$ determines the scalar propagator at zero momentum, which in turn can be used to calculate the mass of the scalar at 1-loop. 
 The path of least resistance is
often to compute the 1-loop potential using functional determinants~\cite{Jackiw:1974cv} and the potential at 2-loops and beyond with prototype Feynman diagrams. 

At 2-loops, no additional tricks are needed and we can compute the effective potential directly from the vacuum diagrams with dressed propagators.
All of the 2-loop diagrams can be computed using two master integrals. The first is a standard one-loop integral
\be
I_1(\Delta) = \int \frac{d^d k}{(2\pi)^d} \frac{1}{k^2 - \Delta + i\e}  = \frac{-i}{(4\pi)^{d/2}} \frac{1}{\Delta^{1-\frac{d}{2}} } \Gamma\left(\frac{2-d}{2} \right)
\ee
The second is the 2-loop scalar vacuum sunset graph with 3 masses~\cite{Davydychev:1992mt, Ford:1992pn, Martin:2001vx}:
\be
I_2(A,B,C) = \int \frac{d^d k_1}{(2\pi)^d} \int \frac{d^d k_2}{(2\pi)^d} 
\frac{1}{\big(k_1^2 - A+ i\e\big)\big(k_2^2 -B+i \e\big) \big( (k_1+k_2)^2 -C+i\e \big)}
\ee
This integral is fully symmetric in $A,B$ and $C$. The full result in $d$ dimensions can be found in~\cite{Davydychev:1992mt} along with its $\e$ expansion. Writing $d=4-2\e$, the expansion is
\begin{multline}
I_2(A,B,C) =\frac{1 }{ (4\pi)^{4-2\e}} \frac{\Gamma(1+\e)^2}{(1-\e)(1-2\e)}\frac{ 1 }{2 C^{2 \e}}\\
\times \Bigg\{-\frac{1}{\e^2} \Big(A+B+C\Big)
+ \frac{2}{\e}\left(A \ln \frac{A}{C} +B \ln \frac{B}{C} \right)  \\
- A \ln^2 \frac{A}{C} - B \ln^2\frac{B}{C} + (C-A-B+C\lambda)\ln\frac{A}{C} \ln \frac{B}{C}\\
-\frac{1}{3}C \lambda \pi^2 - 2 C \lambda \ln \frac{A-B+C - C \lambda}{2C} \ln \frac{B-A+C-C\lambda}{2 C} \\
+2 C \lambda\text{Li}_2\left(\frac{B-A+C-C\lambda}{2 C}\right)+2 C \lambda \text{Li}_2 \left(\frac{A-B+C-C\lambda}{2 C}\right)\Bigg\}
\end{multline}
with
\be
\lambda = \sqrt{\frac{A^2+(B-C)^2 -2 A(B+C)}{C^2}}
\ee
This expansion holds if $\lambda^2 \ge 0$ and $\sqrt{A}+\sqrt{B} \le \sqrt{C}$. The expansion in other
regions can be obtained by permuting $A,B$ and $C$. 

\begin{figure}
\begin{center}
\begin{tabular}{cccc}
A & B & C & D \\[-0.5mm]
\fd{2.5cm}{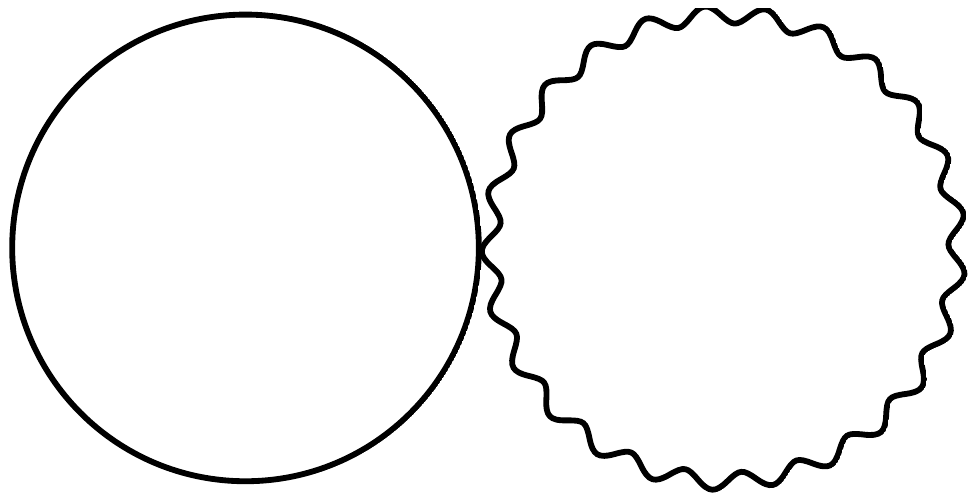} & \fd{1.5cm}{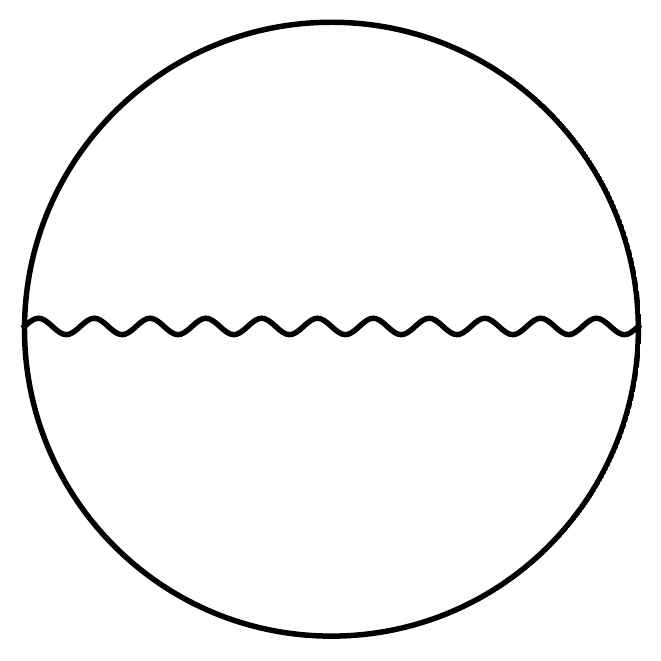} & \fd{1.5cm}{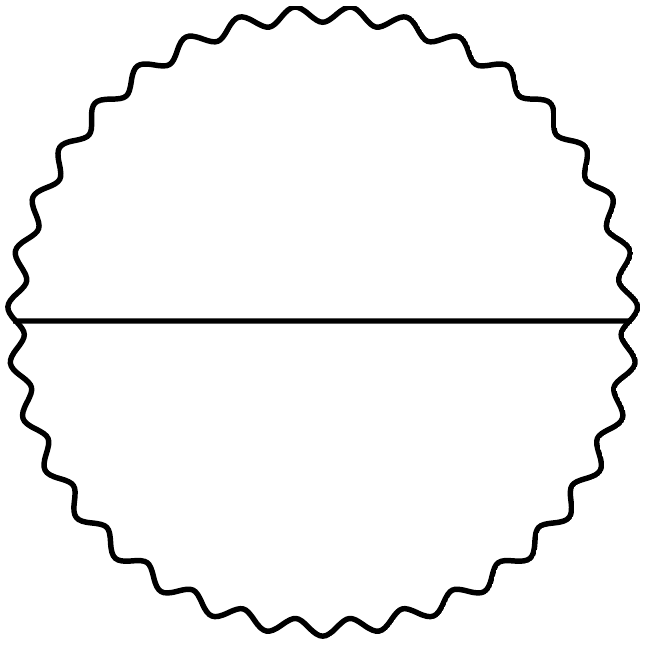} & \fd{1.5cm}{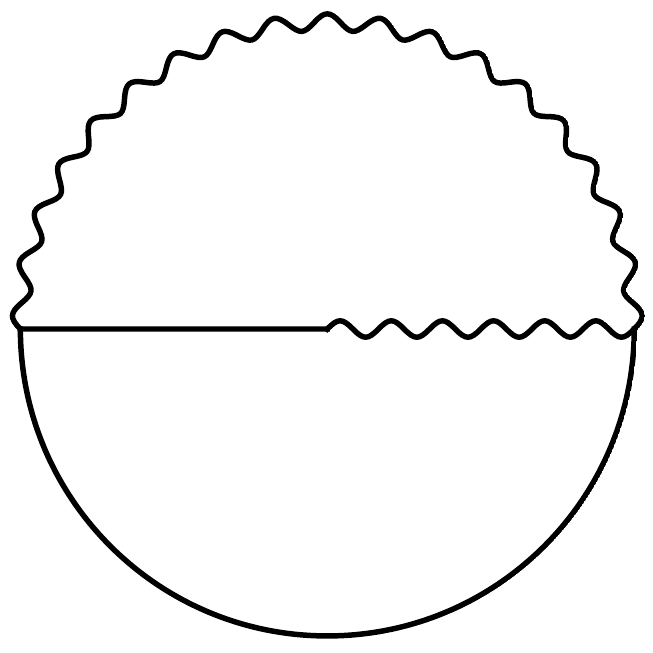}\\[10mm]
E & F & G & H \\[-0.5mm]
\fd{2.5cm}{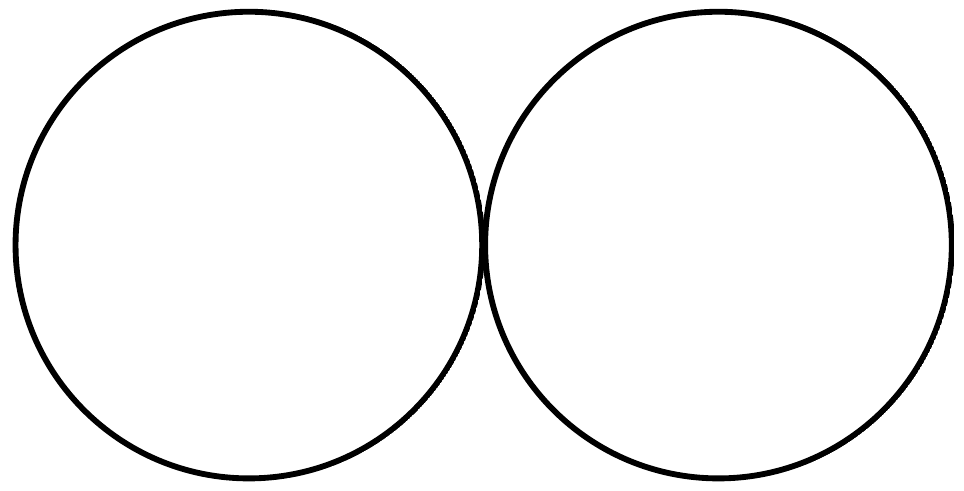} & \fd{1.5cm}{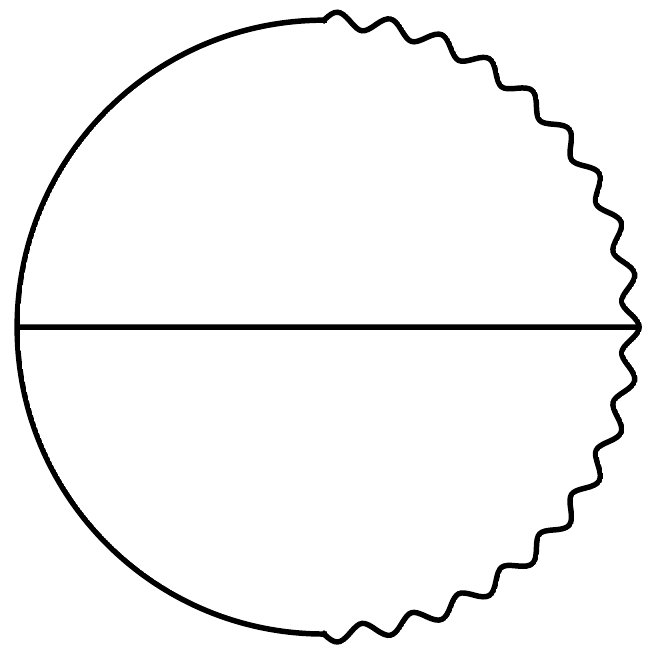} & \fd{1.5cm}{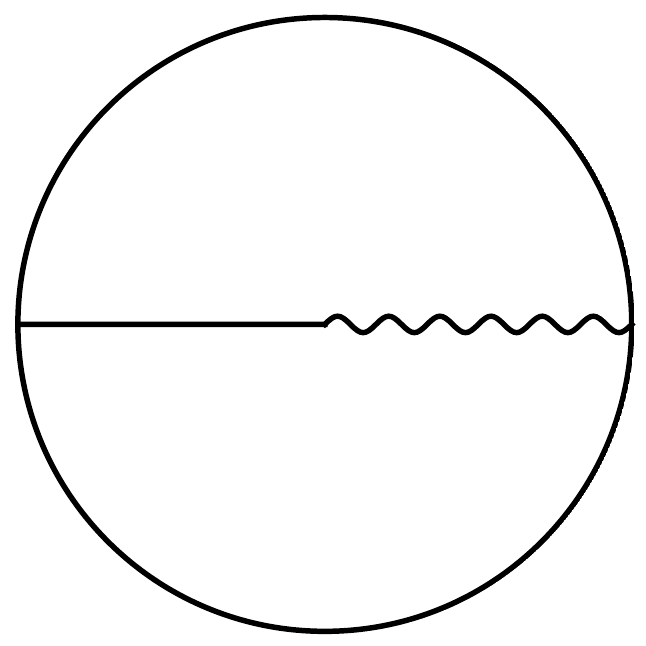} & \fd{1.5cm}{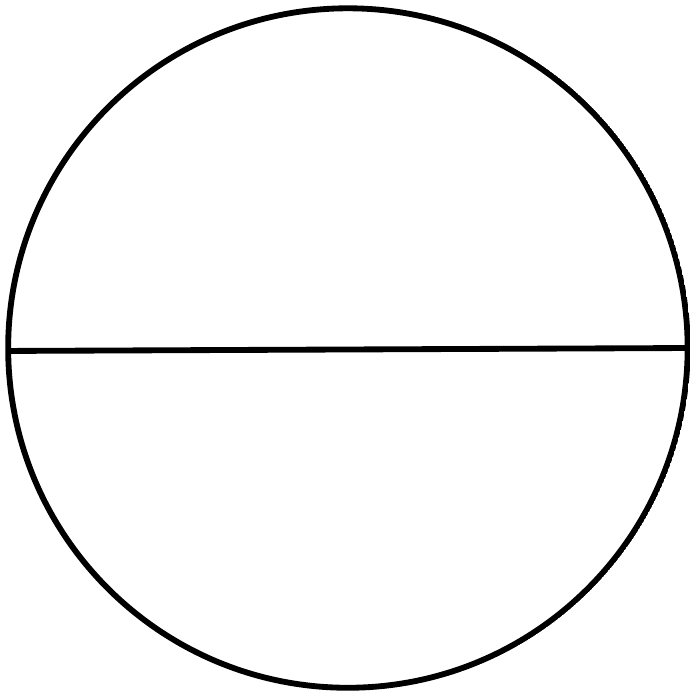}\\[10mm]
I & J & K & L  \\[-0.5mm]
\fd{2.5cm}{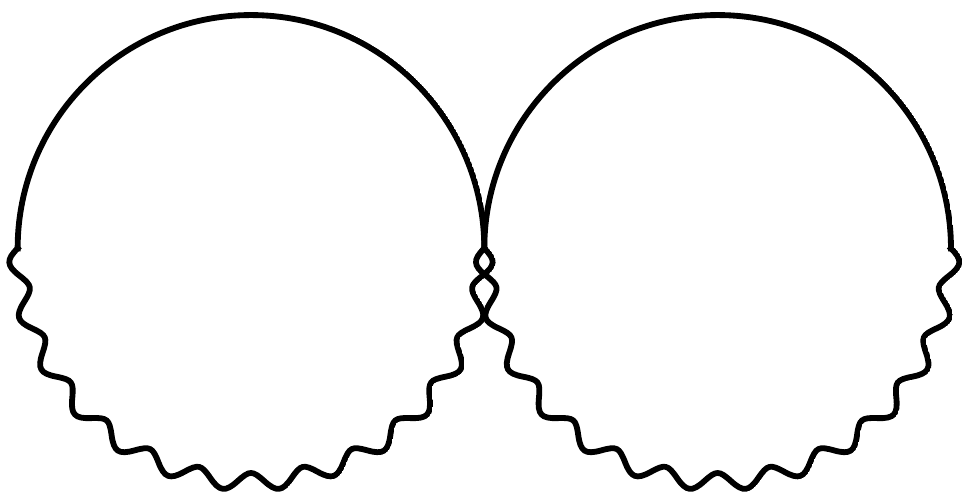} & \fd{1.5cm}{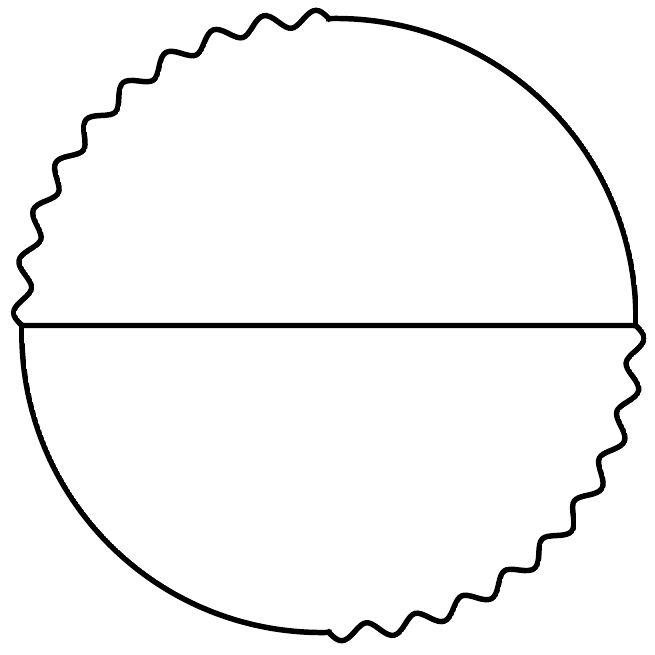} & \fd{1.5cm}{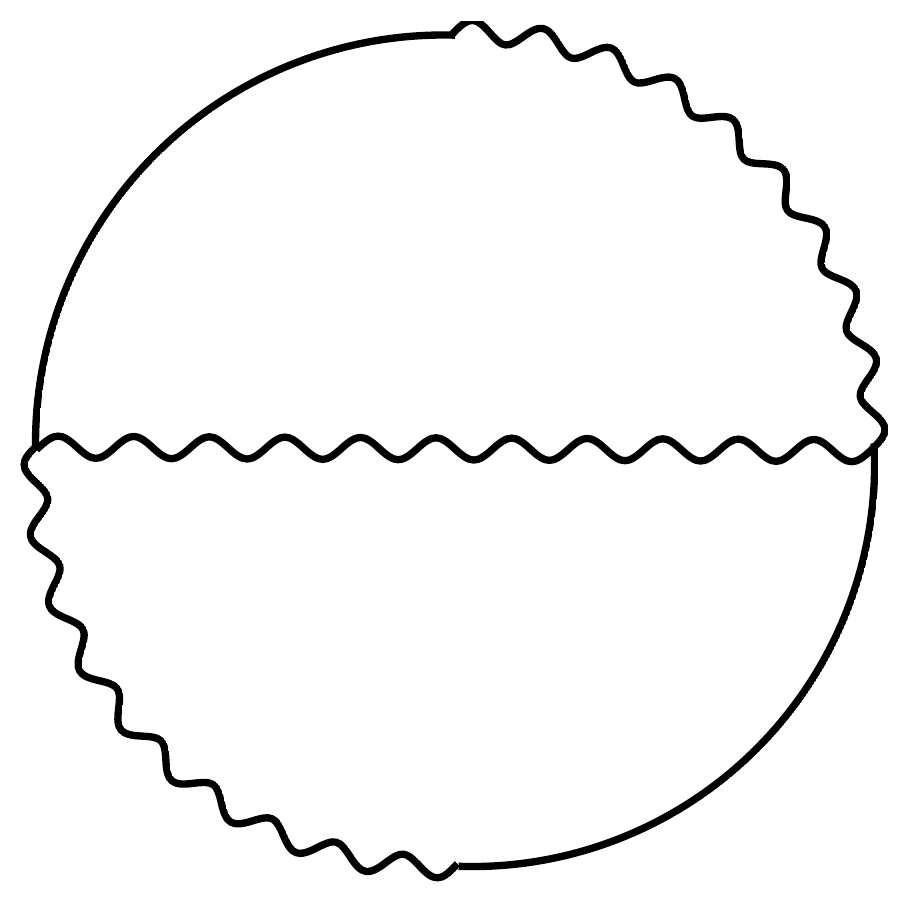} & \fd{1.5cm}{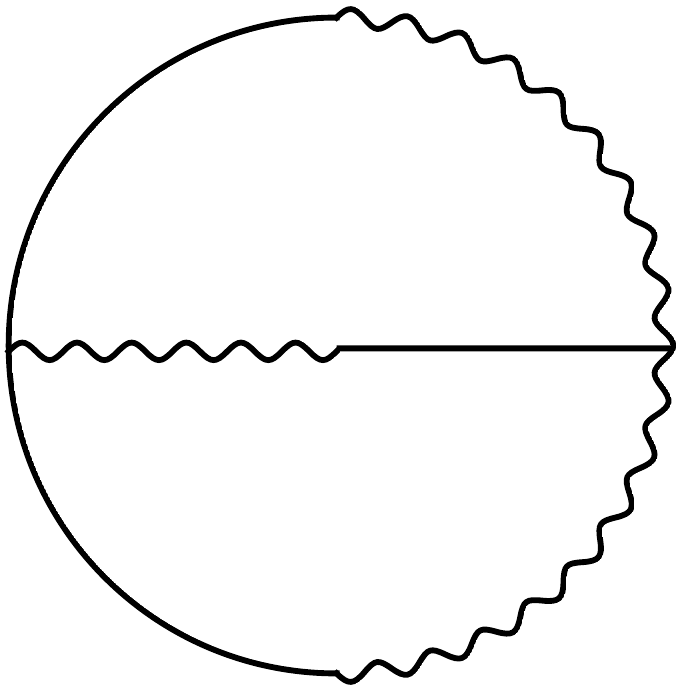}
\end{tabular}
\end{center}
\caption{
There are 12 prototype diagrams which contribute to the 2-loop Coleman-Weinberg potential in $R_\xi$ gauges. Only the 4 diagrams on the top row contribute
at NLO (order $\hbar^2$) when $\lambda \sim \hbar$. 
\label{fig:twoloops}
}
\end{figure}

\newpage
There are 4 diagrams that contribute at order $\hbar^2 e^6$:
\begin{align}
\cM_A &= 
\resizebox{20mm}{!}{
\parbox{30mm}{
\begin{tikzpicture}[]
\node (label) at (0,0)[draw=white]{
       {\fd{3.5cm}{Kb} }
      };
\node[rotate=0,anchor=east] at (-0.6,0.5) {$2$};
\end{tikzpicture}
}}~~
= -\frac{\hbar^2e^2}{2} \int \frac{d^d k_1}{(2\pi)^d}  \int
\frac{d^d k_2}{(2\pi)^d} 
D_{22}(k_1)
\Delta_{\mu\mu}(k_2)  \nn \\
&= \frac{ \hbar^2 \phi^4 e^6}{(16\pi^2)^2}\xi \left[ -12 \ln^2 \frac{e
\phi}{\mu}+ \left(8-3\ln\frac{\lambda \xi}{6 e^2}\right) \ln \frac{e
\phi}{\mu} -\frac{5}{2} - \frac{\pi^2}{16}-\frac{3}{16} \ln^2\frac{\lambda
\xi}{6 e^2} +\ln\frac{\lambda \xi}{6 e^2} \right] +\cdots  \label{VA}\\
\cM_B
 &=
\resizebox{20mm}{!}{
\parbox{30mm}{
\begin{tikzpicture}[]
\node (label) at (0,0)[draw=white]{ 
       {\fd{2.5cm}{Ke}} 
      };
\node[anchor=north] at (0,1.1) {$2$};
\node[anchor=south] at (0,-1.0) {$1$};
\end{tikzpicture}
}}
= \frac{i}{2} e^2
\int \frac{d^d k_1}{(2\pi)^d}  \int \frac{d^d k_2}{(2\pi)^d}
(k_2^\mu - k_1^\mu)(k_1^\nu-k_2^\nu) D_{11}(k_1) D_{22}(k_2) \Delta_{\mu\nu}(k_1+k_2)
\nn\\
&= \frac{ \hbar^2 \phi^4 e^6}{(16\pi^2)^2} \left[(2+6\xi) \ln^2 \frac{e
\phi}{\mu}-  (3+7\xi) \ln \frac{e \phi}{\mu} +\frac{7}{4} +
\frac{\pi^2}{8}+\frac{15}{4}\xi + \frac{3 \pi^2}{8} \xi \right]+\cdots \label{VB}\\
\cM_C
 &=
\resizebox{20mm}{!}{
\parbox{30mm}{
\begin{tikzpicture}[]
\node (label) at (0,0)[draw=white]{ 
       {\fd{2.5cm}{Kf}} 
      };
\node[anchor=south] at (0,0) {$1$};
\end{tikzpicture}
}}
= - i\hbar^2e^4 \phi^2 \int \frac{d^d k_1}{(2\pi)^d}  \int
\frac{d^d k_2}{(2\pi)^d}
\Delta_{\mu\nu}(k_1)\Delta_{\mu\nu}(k_2)D_{11}(k_1+k_2)  \nn \\
&= \frac{ \hbar^2 \phi^4 e^6}{(16\pi^2)^2}\left[(18+6\xi) \ln^2 \frac{e
\phi}{\mu}-  (21+7\xi) \ln \frac{e \phi}{\mu} +\frac{47}{4} +
\frac{7\pi^2}{24}+\frac{15}{4}\xi + \frac{3 \pi^2}{8} \xi \right]+\cdots \label{VC}\\
\cM_D
 &=
\resizebox{20mm}{!}{
\parbox{30mm}{
\begin{tikzpicture}[]
\node (label) at (0,0)[draw=white]{ 
       {\fd{2.5cm}{Kh}}
      };
\node[anchor=south] at (-0.5,0) {$2$};
\node[anchor=south] at (0,-1.0) {$1$};
\end{tikzpicture}
}}
= - 2\hbar^2 e^3 \phi \int \frac{d^d k_1}{(2\pi)^d}  \int \frac{d^d
k_2}{(2\pi)^d} (k_1^\mu - k_2^\mu)\Delta_{\mu\nu}(k_1+k_2)T_{\nu 2}(k_2)
D_{11}(k_1) \nn \\
&=  \frac{ \hbar^2 \phi^4 e^6}{(16\pi^2)^2}\xi  \left[ -12 \ln^2 \frac{e
\phi}{\mu}+ 14 \ln \frac{e \phi}{\mu} -\frac{15}{2}-\frac{3\pi^2}{4}
\right]+\cdots \label{VD}
\end{align}
with the $\cdots$ vanishing as $\lambda \to 0$. 
Kang computed some of the logarithmic terms in these amplitudes in~\cite{Kang:1974yj}, and we agree with his results. 

Adding the contribution of these graphs to the counterterm contribution in Eq.~\eqref{V2ct} gives the $\hbar^2$ contributions to the effective potential:
\begin{multline}
V_2
=  \left(\frac{\hbar}{16\pi^2}\right)^2 e^6 \phi^4
\left[
(10-6\xi) \ln^2\frac{e\phi}{\mu} + \left(-\frac{62}{3} + 4\xi - \frac{3}{2} \xi \ln \frac{\lambda \xi}{6 e^2} \right) \ln \frac{e \phi}{\mu}
\right.\\
\left.
+ \xi\left(-\frac{1}{2} + \frac{1}{4}\ln \frac{\lambda \xi}{6e^2}\right) + \frac{71}{6} \right] + \cdots
\label{V2}
\end{multline}
where again the $\cdots$ vanish as $\lambda \to 0$.

\subsection{RGE cross-check}
As a cross-check, it is easy to verify that $V=V_0+V_1+V_2$ satisfies the renormalization group equation
\be
\Big(\mu \frac{\partial}{\partial \mu}
 - \gamma \phi \frac{\partial}{\partial \phi} 
 + \beta_e \frac{\partial}{\partial e} 
 + \beta_\lambda \frac{\partial}{\partial \lambda}\Big) V= 0 \label{VRGE}
\ee
up to order $\hbar^2$. 
The anomalous dimensions and $\beta$ function coefficients in scalar QED up to 2-loop order
can be extracted from~\cite{Machacek:1983tz,Machacek:1984zw}. We find
\begin{align}
\gamma & = \frac{\hbar}{16\pi^2} e^2(\xi - 3) + \frac{\hbar^2}{(16\pi^2)^2}\left( \frac{10}{3} e^4 +\frac{1}{9} \lambda^2\right) + \cdots\\ 
\beta_\lambda & =
\frac{\hbar}{16\pi^2}\left( 36 e^4-12 e^2 \lambda +\frac{10 \lambda ^2}{3}\right)
+ \frac{\hbar^2}{(16\pi^2)^2}\left(-416 e^6+\frac{316 e^4 \lambda }{3}+\frac{56 e^2 \lambda ^2}{3}-\frac{20 \lambda ^3}{3} \right) +\cdots \nn \\
\beta_e & = 
\frac{\hbar}{16\pi^2} \frac{e^3}{3}  + \cdots \nn 
\end{align}
Using these expansions, our 2-loop effective potential indeed satisfies the RGE.  A similar cross check was done in Landau gauge
for the 2-loop standard model potential in~\cite{Ford:1992pn} and for general renormalizable theories in~\cite{Martin:2001vx}.

In fact, the RGE could have been used to bootstrap all of the $\ln\frac{\phi}{\mu}$ dependence in the 2-loop effective potential. 
Note however that the $\ln \lambda$ and $\ln e$ terms in the 1-loop effective potential are
critical to determining the $\ln\frac{\phi}{\mu}$ dependence at 2 loops, and these terms are not fixed by RG invariance. In other words, we needed the exact 1-loop potential to determine the 2-loop $\ln\frac{\phi}{\mu}$ dependence. 
Coleman and Weinberg argue~\cite[App. A2]{Coleman:1973jx} that one can ignore logarithms of couplings when working in a subtraction scheme with
$\lambda \equiv \frac{d^4 V}{d \phi^4} \Big|_\mu$ for a fixed scale $\mu$.  This may be true, but since the fourth derivative of the potential is not gauge invariant (even at an extremum), this renormalization condition induces unnecessary complications when studying gauge dependence. Moreover, in $\msbar$, logarithms of couplings play an important role and must be kept.

While we are on the subject of RG invariance, note that the $\msbar$ $\beta$ function
coefficients for $e$ and $\lambda$ are gauge invariant. 
This is true to all orders, since the  $\beta$ functions describe the evolution of couplings which can in principle
be measured from  scattering experiments. 
The anomalous dimension for $\phi$ is gauge-dependent;
it describes the evolution of the field strength $Z_\phi$ for $\phi$, which is not itself observable. We will return to the RG when considering the resummed potential in Section~\ref{sec:RG}.

\section{Daisy resummation~\label{sec:daisies}}

\begin{figure}
\begin{center}
\fd{3cm}{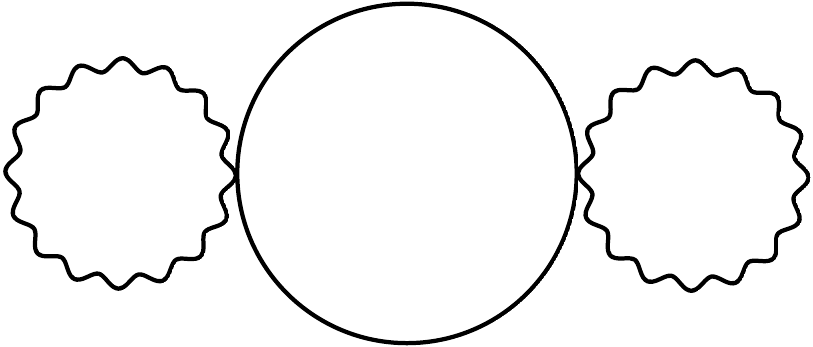}~~\fd{3cm}{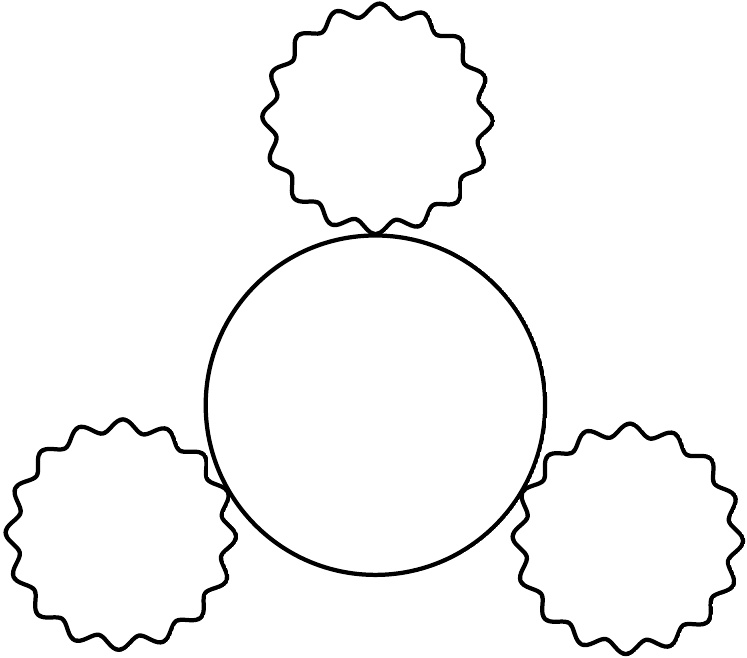}~~\fd{3cm}{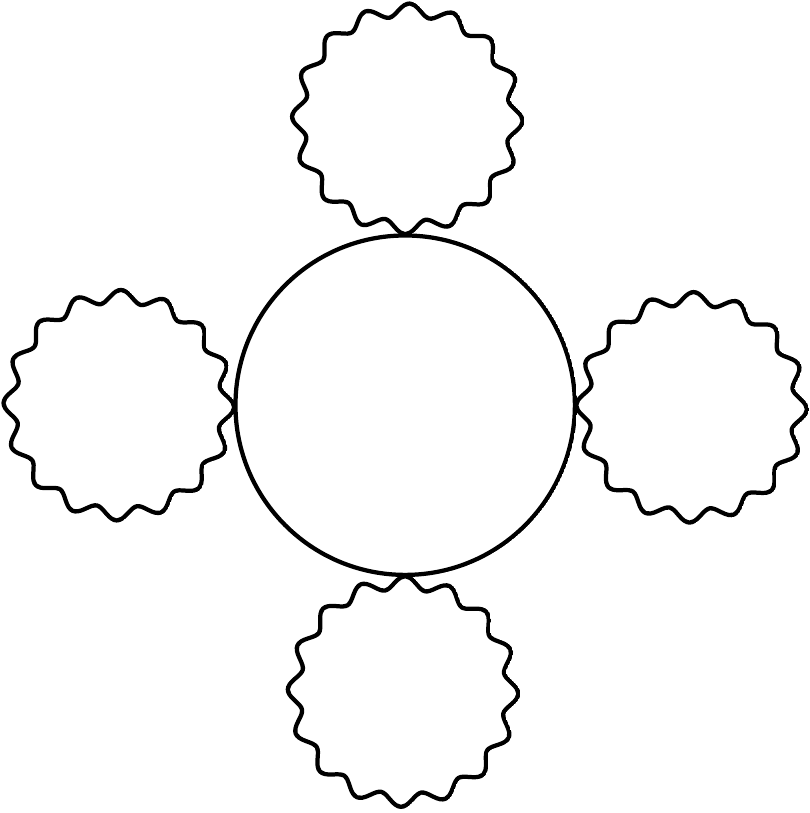}
\end{center}
\caption{
Example daisy graphs which contribute inverse powers of $\lambda$. 
\label{fig:daisies}
}
\end{figure}

As pointed out by Nielsen~\cite{Nielsen:1975fs,Nielsen:1987ht}, there can be contributions to the effective potential proportional to $\lambda^{-1} e^{10}$ coming from {\it daisy diagrams}, like those in Figure~\ref{fig:daisies}. When $\lambda \sim e^4$,
these terms are of the same order as the $e^2 \lambda$ terms in the 1-loop potential and the $e^6$ terms in the 2-loop potential. Thus they should be relevant to showing gauge invariance of physical quantities
at order $e^6$ in this power counting.  Nielsen argued that the reason the scalar mass Kang calculated does not satisfy his identity was due to the absence of these terms~\cite{Nielsen:1975fs}. 
Some time later, Johnston~\cite{Johnston:1986cp} showed how these terms can be summed into a dressed propagator for the scalar fields, suggesting that these terms could be computed and the Nielsen identity restored,
although no explicit contributions to the potential were provided.
Shortly afterwards Bazeia~\cite{Bazeia:1988pz} showed that even at 1-loop, where daisies are irrelevant, the vacuum energy in the Coleman-Weinberg model has gauge-dependence. 
In this section, we compute all of the daisy graphs relevant at $e^6$ (and some of the subleading daisy graphs as well to demonstrate their relevance in Landau gauge). In the next section 
we demonstrate that, after carefully keeping track of the independent variables, the effective potential at its minimum is indeed gauge-invariant.

Before beginning, it is worth remarking that the daisy resummation we perform here is related to, but not identical to, daisy resummation in finite-temperature field theory~\cite{Kirzhnits:1972ut,Weinberg:1974hy,Carrington:1991hz}. At finite temperature, the resummation of daisy diagrams is necessary to calculate a critical temperature because, in the limit $m\ll T$, new infrared divergences arise. Daisy resummation as a solution to infrared problems associated with massless
Goldstone bosons has recently been proposed in~\cite{Martin:2014bca,Elias-Miro:2014pca}.
The relevance of daisy resummation to solving gauge-dependence problems at finite temperature has also been discussed~\cite{Patel:2011th,Wainwright:2011qy}. 
It is therefore not surprising that daisy resummation is relevant to the gauge-dependence problem in the Coleman-Weinberg model. Nevertheless, the relevant calculations have never been done, to our knowledge, which is why we resum the relevant daisy graphs here.

Normally, one does not get inverse powers of coupling constants from Feynman diagrams. Indeed, at any fixed order in perturbation theory, there are always positive powers of couplings, even in effective potential
calculations. However, the effective potential always involves summing an infinite number of graphs, namely those with an arbitrary number of background field insertions. It is this infinite sum
which gives the $\ln\phi$ dependence in the effective potential and which can generate inverse powers of couplings. As discussed in Section~\ref{sec:twoloopCW}, we simplify the infinite sums by
using dressed propagators. For example, we see from Eq.~\eqref{D11def} that the $\phi_1$ propagator is $D_{11} = \frac{i}{k^2 - \frac{\lambda}{2} \phi^2}$
which has an effective mass $m^2 = \frac{\lambda}{2} \phi^2$. In the daisy graphs, each photon loop (the petals) gives a factor proportional to $e^2 \phi^2$, each vertex gives a factor of $e^2$, and the loop integral over the scalar propagators
can give inverse powers of the effective mass. For example, a 4-loop 3-petal daisy will give
\be
\resizebox{20mm}{!}{
\parbox{30mm}{
\begin{tikzpicture}[]
\node (label) at (0,0)[draw=white]{ 
       {\fd{5cm}{Daisy3}} 
      };
\node[anchor=south] at (-0.6,-0.35) {$1$};
\node[anchor=south] at (0.6,-0.35) {$1$};
\node[anchor=south] at (0,-1.55) {$1$};
\end{tikzpicture}
}}
~~~~
\propto (e^2)^3(e^2\phi^2)^3\int \frac{d^4 k}{2\pi^4}\left( \frac{i}{k^2 - \frac{\lambda}{2} \phi^2} \right)^3 \propto \phi^4 \frac{e^{12}}{\lambda}
\ee
Here, we have simply done the integral by dimensional analysis, since it is UV and IR finite. This graph therefore contributes at order $e^8$ in the $\lambda \sim e^4$ power-counting. It is therefore beyond the order we need for the first non-trivial gauge-invariance check of $\Vmin$ (not to mention that this particular loop is itself $\xi$-independent).

It is not hard to see, using dimensional analysis, that the only graphs which could contribute at order $e^6$ (with the $\lambda \sim e^4$ power-counting) must involve $\phi_2$ propagators and have petal-type
photon loops. The petals factorize off from central disk, and from each other. Thus a daisy graph with $n$ petals has the form
\be
I_n = i^{n+1}\frac{e^{2n}}{2n} A_{n \phi_2} (A_\gamma)^n
\ee
where $A_\gamma$ includes the photon loop and its counterterm, $A_{n\phi_2}$ is the scalar integral with $n$ $\phi_2$ propagators, and 
the factor $\frac{1}{2n} e^{2n}$ comes from the vertex Feynman rule and the symmetry factor for the circle. 

Each photon loop gives
\begin{align}
  A_{\gamma}^{\text{loop}} &= {\fd{1.5cm}{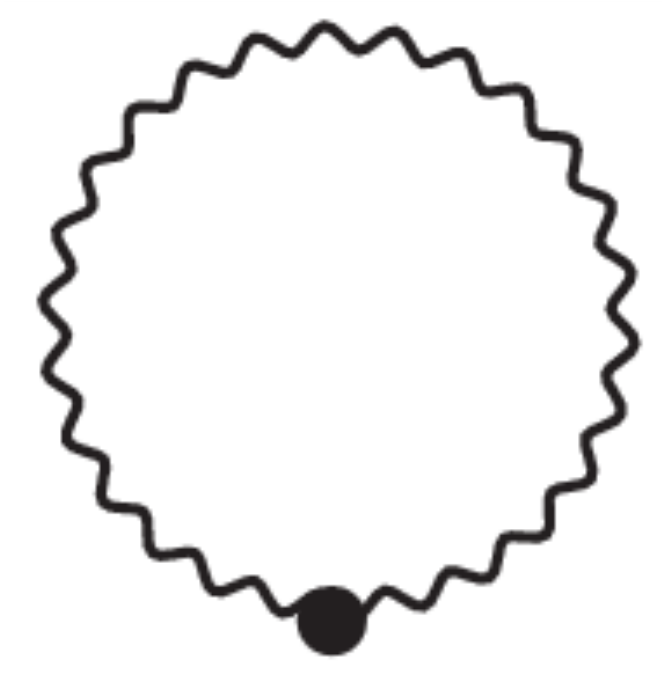}} = \int \frac{d^d k}{( 2 \pi )^d} \Delta_{\mu \mu} \nn \\
&= 
  \frac{\hbar}{16 \pi^2}e^2 \phi^2 \left[ \frac{3}{\e} + \left( 1 - 6 \ln
  \frac{\phi e}{\mu}  \right) + \e \left( 1 + \frac{\pi^2}{4} - 2 \ln
  \frac{\phi e}{\mu} + 6 \ln^2 \frac{\phi e}{\mu} \right) + \cdots \right] 
\end{align}
The UV pole in this loop is removed by the 1-loop counterterm:
\be
A_{\gamma}^{\text{c.t.}} =   \frac{\hbar}{16 \pi^2}e^2 \phi^2 \left[- \frac{3}{\e}\right]
\ee
 The $\cO(\e)$ in $A_\gamma^{\text{loop}}$ can be important if the scalar loop it multiplies is UV-divergent. This only happens for two petals ($n=2$). For $n=2$, the scalar loop gives
\be
A_{2 \phi_2} 
=
\resizebox{20mm}{!}{
\parbox{30mm}{
\begin{tikzpicture}[]
\node (label) at (0,0)[draw=white]{ 
       {\fd{2.5cm}{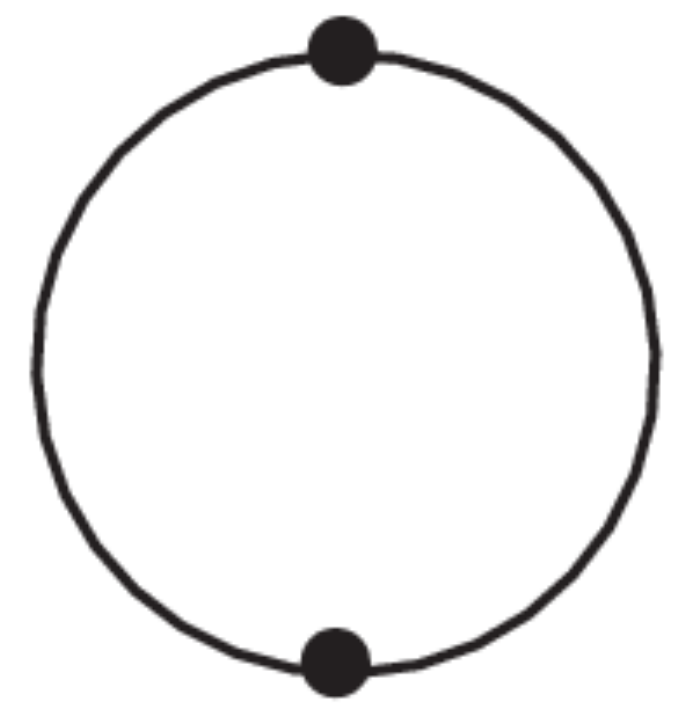}} 
      };
\node[anchor=south] at (-0.7,-0.25) {$2$};
\node[anchor=south] at (0.7,-0.25) {$2$};
\end{tikzpicture}
}}
=  \int \frac{d^d k}{( 2 \pi )^d} D_{22}^2 =- i \frac{\hbar}{16
  \pi^2} \left[ \frac{1}{\varepsilon} + \frac{3 e^2 \xi}{\lambda} + \cdots
\right]
\ee
Since the $\frac{1}{\e}$ term in $A_{2 \phi_2}$ has no $\frac{1}{\lambda}$ piece, the cross term between it and the $\cO(\e)$ part of $A_\gamma^{\text{loop}}$ will not contribute at order $e^6$ (with $\lambda\sim e^4$). Thus, for all the daisies, we can drop the $\cO(\e)$
terms in $A_\gamma^{\text{loop}}$ and take
\be
A_\gamma = A_{\gamma}^{\text{loop}} +A_{\gamma}^{\text{c.t.}}  =  \frac{\hbar }{16 \pi^2}e^2 \phi^2 \left( 1 - 6 \ln
  \frac{e \phi}{\mu}  \right) 
\ee

For $n>3$, the scalar loop is UV and IR finite. We find
\be
A_{n \phi_2}  =\fd{1.5cm}{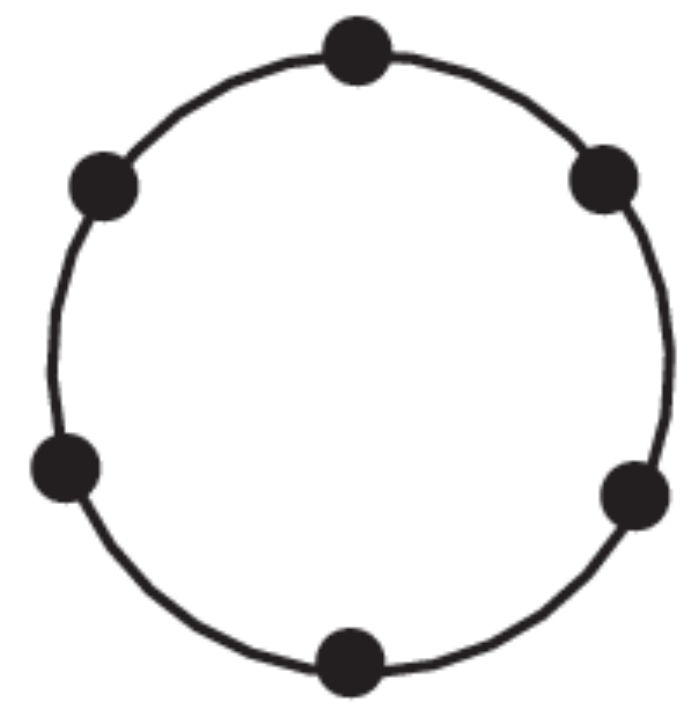} = \int \frac{d^4 k}{( 2 \pi )^4} D_{22}^n 
=\frac{i\hbar}{16\pi^2}  \phi^4 \frac{e^2 \lambda \xi}{12(n-1)}\left(\frac{-6 i }{\lambda \phi^2} \right)^n
\ee
so that
\be
  I_n = \frac{\hbar}{16 \pi^2} \left( - \frac{1}{24} e^2 \lambda \xi \phi^4
  \right) \frac{1}{ n ( n - 1 )} \left[ \frac{\hbar e^4}{\left( 16
  \pi^2 \right) \lambda} \left( 6 - 36 \ln \frac{e \phi }{\mu}  \right)
  \right]^n
\ee
Each term in this series contributes at order $e^6$ when $\lambda \sim e^4$. Thus they are all equally important for checking gauge invariance and we must sum the series. Summing the series is easy enough to do using
\begin{equation}
  \sum_{n = 2}^{\infty} \frac{1}{n ( n - 1 )} x^n = x + ( 1 - x ) \ln ( 1 - x )
\end{equation}
which gives
\be
V^{e^6\text{daisies}} = \phi^4 \frac{\hbar}{16\pi^2}\left(- \frac{e^2 \lambda  \xi}{24}\right)
\left[\frac{\lhat(\phi)}{\lambda} + \big(1 -\frac{ \lhat(\phi)}{\lambda} \big) \ln \big( 1 - \frac{\lhat(\phi)}{\lambda} \big) \right]
\label{daisyfull}
\ee
where
\be
\lhat(\phi) \equiv \frac{\hbar e^4}{ 16 \pi^2 } \left(6-36 \ln \frac{e \phi}{\mu}  \right) \label{lhatdef}
\ee
We have defined $\lhat(\phi)$ so that   according to  Eq.~\eqref{lambdacond} $\lhat(v) = \lambda$ at the scale $v$ where the 1-loop potential has its minimum. 
Remarkably, while each daisy graph with $n>1$ is individually power-divergent as $\lambda \to 0$ with $e$ fixed, the sum of all daisies scales
only like $\ln \lambda$.

Before moving on, it is worth pointing out that daisy resummation is important even in  Landau gauge, $\xi=0$. In Landau gauge, there is no kinetic mixing and the scalar propagators are 
\be
D_{11} = \frac{i}{k^2 - \frac{\lambda}{2} \phi^2},\quad D_{22} = \frac{i}{k^2 - \frac{\lambda}{6} \phi^2},\quad 
\ee
These propagators still have $\lambda$-dependent masses and can produce $\frac{1}{\lambda}$ dependence from daisy graphs. For example, summing the daisy graphs with $\phi_1$ or $\phi_2$ running in the loop gives
\be
V^{e^8\text{daisies}} 
=-\frac{1}{2} \phi^4 \frac{\hbar}{16\pi^2}\left[  \frac{\lambda^2}{4} F\big(\frac{2\lhat}{\lambda}\big)+\frac{\lambda^2}{36} F\big(\frac{6\lhat}{\lambda}\big)\right]
\ee
where
\be
F(x) = \frac{x^2-4x}{48}-\frac{1}{2}(1-\frac{x}{6})^2  \ln \left(1-\frac{x}{6}\right)
\ee
These terms are important, but since $\lambda \sim e^4$, they have effects comparable to terms in the 3-loop Coleman-Weinberg potential. Thus,
an advantage of Landau gauge is that it postpones the relevance of daisy resummation by one loop. Landau gauge does not however let us ignore
the daisy graphs completely. 

Two recent papers also observed that resummation of
certain graphs to all orders is necessary starting at 3 loops in Landau gauge~\cite{Martin:2014bca,Elias-Miro:2014pca}.
 These two papers are concerned with resolving an infrared divergence problem associated with massless Goldstone bosons
starting at 3 loops. While these two papers discuss diagrams similar to the ones here,
the problem they solve is different (infrared divergences, not gauge-dependence) and their results are not directly transferable. However, these two papers, along with the earlier work in~\cite{Johnston:1986cp,Quiros:1992ez}, 
do explain in a more systematic way how daisy and other relevant diagrams can be resummed through a modification of the effective propagators.

In summary, the full Coleman-Weinberg potential up to order $e^6$ with $\lambda \sim e^4$ is
the sum of Eqs.~\eqref{V0}, \eqref{V1}, \eqref{V2} and \eqref{daisyfull}. It is helpful to write the result as
\be
V = \VLO + \VNLO + \cdots
\ee
where the leading-order (LO) potential
\be
 \VLO= \frac{\lambda}{24} \phi^4 + \frac{\hbar e^4}{16\pi^2} \phi^4  \left(-\frac{5}{8} + \frac{3}{2} \ln \frac{e \phi}{\mu} \right)
\label{VLO}
\ee
scales as $\cO(\hbar)$ when $\lambda \sim \hbar e^4$ and the next-to-leading order (NLO) potential, scaling like $\cO(\hbar^2)$, is
\begin{empheq}[box=\fbox]{multline}
\\[-4mm]
\hspace{-80mm}
 \VNLO= \frac{\hbar e^2 \lambda }{16\pi^2} \phi^4  \left(\frac{\xi}{8} - \frac{\xi}{24} \ln \frac{e^2 \lambda \xi \phi^4}{6 \mu^4} \right) \\
\hspace{-10mm}
+ \frac{\hbar^2 e^6}{(16\pi^2)^2} \phi^4
\left[
(10-6\xi) \ln^2\frac{e\phi}{\mu} + \left(-\frac{62}{3} + 4\xi - \frac{3}{2} \xi \ln \frac{\lambda \xi}{6 e^2} \right) \ln \frac{e \phi}{\mu}
+ \xi\left(-\frac{1}{2} + \frac{1}{4}\ln \frac{\lambda \xi}{6e^2}\right) + \frac{71}{6}\right]
~~~~~~~~~~~
\\
+{
  \phi^4 \frac{\hbar e^2 \lambda}{16\pi^2}\left(- \frac{\xi}{24}\right)
 \left[\frac{\lhat(\phi)}{\lambda} + \big(1 -\frac{ \lhat(\phi)}{\lambda} \big) \ln \big( 1 - \frac{\lhat(\phi)}{\lambda} \big) \right]
} 
\\[2mm]
\label{V2boxed}
\end{empheq}
with $\lhat(\phi)$ is defined in Eq.~\eqref{lhatdef}. Note that there are tree and 1-loop contributions to the LO potential and that the NLO potential get contributions
from 1-, 2- and all higher order loops. 

\section{Gauge invariance of $\Vmin$ \label{sec:GIphys}}
With the 2-loop Coleman-Weinberg potential in hand and the contribution of daisy graphs understood, we can now demonstrate
gauge-invariance of the potential at its minimum. 

Recall from Eq.~\eqref{Vmin1} that at leading order (LO), $\Vmin^{\text{LO}} = -v^4e^4 \frac{3}{128\pi^2}$. One might naturally expect that the next-to-leading order (NLO)
contribution could be written as $\Vmin^{\text{NLO}} = v^4 e^6 C$ for some $\xi$-independent constant $C$. Unfortunately, we cannot expect $\Vmin$ to be explicitly $\xi$-independent when written
in terms of $v$. The problem is that $v=\langle\phi\rangle$ is a field value and therefore gauge-dependent, so infects all dimensionful quantities expressed in terms of it. One alternative is
to calculate not $\Vmin$ but the ratio of $\Vmin$ to some other dimension-four quantity, such as $m_S^4$, where $m_S$ is the scalar mass. Indeed, Coleman and Weinberg~\cite{Coleman:1973jx} and later Kang~\cite{Kang:1974yj} discussed 
gauge invariance of the dimensionless ratio of the scalar to vector masses.  However, we would really like to see that $\Vmin$ is physical on its own. But then, if not $v$, what are we to  express $\Vmin$ in terms of? 

An alternative to expressing $\Vmin$ in terms of $v$ is to express it in terms of the renormalization group scale $\mu$. This scale is as physical as the $\msbar$ couplings: the two are intrinsically connected. 
To be concrete, let us define the scale $\muX$ as the scale where Eq.~\eqref{lambdacond2} is satisfied exactly. That is, $\muX$ is defined by the exact relation
\be
\lambda(\muX) \equiv \frac{\hbar}{16\pi^2} e^4(\muX)\Big\{6-36 \ln [e(\muX) ]\Big\} \label{lamerel}
\ee
Since this relation is exact, we no longer can or need to solve for $e^6$ terms in the relation between  $\lambda$ and $e$, as was done in~\cite{Kang:1974yj}. One can instead now solve for corrections to $v$
\be
v= \langle \phi \rangle = \muX + v_1 + v_2 + \cdots
\ee
with $v_1 \sim e(\muX)^2$, $v_2 \sim e(\muX)^4$, etc., and $\muX$ is defined by Eq.~\eqref{lamerel}. 
Then one can consistently expand $\Vmin$ at $\mu=\muX$:
\begin{multline}
\Vmin = \VLO(\muX) \\[-5mm]
+  \VNLO(\muX) + v_1 \overbrace{\frac{d}{d\phi} \VLO\Big|_{\phi=\muX}}^{=0} \hspace{6cm}\\
+ V^{\text{NNLO}}(\muX) + v_1  \frac{d}{d\phi} \VNLO\Big|_{\phi=\muX}+
v_2  \underbrace{  \frac{d}{d\phi} \VLO\Big|_{\phi=\muX}}_{=0}
 + \frac{1}{2} v_1^2   \frac{d^2}{d\phi^2} \VLO\Big|_{\phi=\muX} + \cdots 
 \\[-7mm]
 \label{Vnewis}
\end{multline}
where $\VLO$ is the effective potential truncated to order $\hbar$ with $\lambda \sim \hbar$, $\VNLO$ is truncated to $\hbar^2$ and so on,
as in Eqs.~\eqref{VLO} and \eqref{V2boxed}.

A convenient feature of setting $\phi=\muX$ is that the daisy contribution simplifies. Indeed, from Eq.~\eqref{lhatdef} we see that when $\phi=\mu=\muX$ then $\lhat(\muX) = \lambda(\muX)$. Then
the entire daisy contribution on the last line of Eq.~\eqref{V2boxed} reduces to simply
\be
V^{\text{NLO, daisies}} = -\frac{\xi}{24}\phi^4 \frac{\hbar e^2 \lambda}{16\pi^2} = -\frac{\xi}{24}\frac{\hbar^2 e^6}{(16\pi^2)^2} (6-36 \ln e)
\ee
Also, as indicated in Eq.~\eqref{Vnewis}, we can use that $V^{\text{LO}\, \prime}(\muX)=0$, which was the defining equation for $\muX$. This simplifies $\Vmin$ to
\be
\Vmin = \VLO(\muX) +  \VNLO(\muX) +\cdots
\ee
Plugging in Eqs.~\eqref{VLO} and \eqref{V2boxed} and using Eq.~\eqref{lamerel} we then find
\be
\Vmin = - \frac{ 3\hbar e^4}{128 \pi^2} \muX^4 + \frac{e^6 \hbar^2}{( 16 \pi^2 )^2} \muX^4 \left(
  \frac{71}{6}  - \frac{62}{3}\ln e+ 10  \ln^2 e \right)
    \label{Vminofmu}
\ee
which is manifestly gauge-invariant! The daisies have exactly canceled the $\xi$ dependence of the NLO 1-loop and 2-loop contributions.

Next, let us look at a field value expressed in terms of the $\msbar$ scale $\mu$, to double check that somehow all of its gauge-dependence is not miraculously absent.
Consider the value of the field where the potential is zero $\Lambda_I$, which in the Standard Model is sometimes
given an interpretation as an instability scale~\cite{Buttazzo:2013uya}. Setting $V(\Lambda_I)=0$ gives a different relation between $\lambda$ and $e$ than $V'(v)=0$ did. The condition on the running couplings so that $\VLO=0$ at $\phi=\mu=\muI$ is
\be
\lambda(\muI) = \frac{ \hbar}{16 \pi^2}e(\muI)^4 \Big\{15 - 36 \ln [e(\muI)]  \Big\}
\ee
At NLO, 
we then find
\be
\Lambda_I = \muI - \frac{\VNLO(\muI)}{V^{\text{LO}\,\prime}(\muI)} + \cdots = \muI - \frac{32\pi^2}{3 e^4 \hbar \muI^3} \VNLO(\muI)
\ee
To evaluate the daisy contribution, we can no longer use $\lhat=\lambda$. Instead we now find
\be
\frac{\lhat}{\lambda} = \frac{\lhat(\phi=\mu=\muI)}{\lambda(\mu=\muI)} =1- \frac{3}{5-12\ln e} 
\ee
Up to NLO, we then find
\begin{multline}
\Lambda_I = \muI - \frac{\VNLO(\muI)}{V^{\text{LO}\,\prime}(\muI)} + \cdots\\
 = \muI 
+\muI \frac{\hbar e^2}{16\pi^2}
\left\{
-\frac{71}{9}-\frac{11 \xi }{12}
+\left(
\frac{5 \xi}{2}+\frac{124}{9}\right) \ln e-\frac{20}{3} \ln^2e 
+\frac{1}{4} \xi  \ln \left[\frac{h \xi  }{32 \pi ^2}(5-12 \ln e) \right]
\right.\\
\left.
-\frac{\xi}{12}\left(-2 + 12 \ln e -3 \ln\frac{3}{5-12 \ln e}\right)
\right\} + \cdots
\label{LambdaIofxi}
\end{multline}
This instability scale is linearly dependent on the gauge-parameter $\xi$, and therefore should not be used to draw physical conclusions. The $\xi$ dependence of other field values can be computed in a similar way, confirming that they are indeed unphysical.

\section{Renormalization group improvement~\label{sec:RG}}
At this point we have shown that if the effective potential in scalar QED is expressed in terms of the $\msbar$ couplings $e$ and $\lambda$ and the scale $\muX$ where they satisfy Eq.~\eqref{lamerel},
then the value of the potential at the quantum minimum $\Vmin$ is gauge invariant. We checked this to the first non-leading order, which required the 2-loop potential and the summation of the leading daisy
diagrams. A natural question is whether we {\it must} express $\Vmin$ in terms of $\muX$?
Clearly, we should be able to write any physical quantity in terms of any other scale. More importantly, the check required the truncation of the effective potential to order $e^6$ with 
$\lambda(\muX) \sim e(\muX)^4$. We would like to be able to use effective potentials in other contexts, where the couplings are defined at some other arbitrary scale $\muY$ and the renormalization
group is used to evolve the potential to a scale near its minimum. So how are we to use this unusual truncation in a RG-improved effective potential, where some $\lambda$ and $e$ dependence
is necessarily included to all orders in $\hbar$? In this section, we show how to answer both these questions, and that the answers are related.

\subsection{Calculation-scale invariance \label{sec:CSI}}
First, let us consider how the calculation of $\Vmin$ would change if we had calculated the effective potential with the couplings defined at a scale $\muY$ instead of $\muX$. 
At $\muY$, the potential obviously has the same form, Eq.~\eqref{V2boxed}, but with $e$ and $\lambda$ interpreted as $e_\Y \equiv  e(\muY)$ and $\lambda_\Y \equiv \lambda(\muY)$ rather than $e_\X \equiv e(\muX)$ and $\lambda_\X \equiv \lambda(\muX)$. That is,
\be
V= \frac{\lambda_\Y}{24} \phi^4 
+ \frac{\hbar}{16\pi^2}\phi^4
\left[
e_\Y^4\left(-\frac{5}{8} + \frac{3}{2} \ln \frac{e_\Y \phi}{\muY} \right)
+e_\Y^2 \lambda_\Y \xi \left(\frac{1}{8} - \frac{1}{24} \ln \frac{e_\Y^2 \lambda_\Y \xi \phi^4}{6 \muY^4} \right) \right] + \cdots
\label{VY}
\ee
For simplicity, let us assume for now that $\muY$ is close enough to $\muX$ so that $\lambda_Y$ is still much smaller than $e_Y^2$.
For the daisy contribution evaluated at $\muY$, we can then still use $\lhat \sim \lambda$, since the difference is higher order.
This lets us continue to drop most of Eq.~\eqref{daisyfull}. 

The easiest way to study $V$ with couplings at $\muY$ is simply to use the renormalization group to move the couplings back to $\muX$. Then we can recycle the previous analysis. Expanding
Eq.~\eqref{e2mu} gives
\be
e_\Y = e_\X  + \frac{\hbar}{48\pi^2} e_\X^3 \ln\frac{\muY}{\muX} + \cdots
\ee
Expanding Eq.~\eqref{lambdamu} gives
\be
\lambda_\Y = \lambda_\X +\frac{\hbar}{16\pi^2} \left[ 36 e^4_\X - 12 e_\X^2 \lambda_\X\right] \ln\frac{\muY}{\muX}+ \frac{\hbar^2}{(16\pi^2)^2}e^6_\X\left[-416 \ln\frac{\muY}{\muX}-192\ln^2\frac{\muY}{\muX} \right] + \cdots
\ee
Plugging these into Eq.~\eqref{VY} we find
\begin{multline}
V= \frac{\lambda_\X}{24} \phi^4 \\
+ \frac{\hbar}{16\pi^2} \phi^4
\left[
e_\X^4\left(-\frac{5}{8} + \frac{3}{2} \ln \frac{e_\X \phi}{\muX} \right)
+e_\X^2 \lambda_\X 
 \left(\frac{\xi}{8} - \frac{\xi}{24} \ln \frac{e_\X^2 \lambda_\X \xi \phi^4}{6 \muX^4}+ \frac{\xi-3}{6}\ln\frac{\muY}{\muX} \right) \right] +\cdots
\label{VYtoX}
\end{multline}
We have not even shown the 2-loop terms, but already one can see that this potential is not identical to what we got from computing $V$ starting at $\mu=\muX$:
there is an uncanceled $\ln\frac{\muY}{\muX}$ term.  Despite this modification of the potential, we still find that
\be
\Vmin = \frac{e^4 \hbar}{16 \pi^2} \muX^4 \left( - \frac{3}{8}
  \right) + \frac{e^6 \hbar^2}{( 16 \pi^2 )^2} \muX^4 \left(
  \frac{71}{6}  - \frac{62}{3}\ln e+ 10  \ln^2 e \right) + \cdots
\ee
exactly as in Eq.~\eqref{Vminofmu}. That is, the new $\ln\frac{\muY}{\muX}$ term has no effect on the value of the potential at the minimum.

That the $\muY$ dependence would drop out of $\Vmin$ was anticipated in Section~\ref{sec:rescale}, with $\mu_0$ replaced by $\muY$ and $\mu$ replaced by $\muX$.
 We next review and extend that discussion in the context of this concrete example.

\subsection{Discussion \label{sec:discussion}}

There is a quick way to see why computing the potential at a different scale affected $V(\phi)$ but not $\Vmin$. The effective potential satisfies the RGE in Eq.~\eqref{VRGE}:
\be
\Big(\mu \frac{\partial}{\partial \mu}
 - \gamma \phi \frac{\partial}{\partial \phi} 
 + \beta_e \frac{\partial}{\partial e} 
 + \beta_\lambda \frac{\partial}{\partial \lambda}\Big) V= 0 
\ee
This equation says that the explicit $\mu$-dependence in the effective potential is exactly compensated for by a rescaling of the coupling constants, according to their $\beta$-functions, and a rescaling of the field
$\phi$, according to its anomalous dimension. In writing the potential computed at $\muY$ in terms of couplings at $\muX$ we only rescaled the couplings, not the field. Indeed, the extra $\frac{\xi-3}{6} \ln\frac{\muY}{\muX}$ term in Eq.~\eqref{VYtoX} would be exactly removed if we rescaled the field as $\phi \to \kappa \phi$ with
\be
\kappa = 1 + \frac{e^2}{16\pi^2}(3-\xi)\ln\frac{\muY}{\muX} + \cdots
\ee
Because the two potentials (calculated at $\muX$ or $\muY$) differ by terms which can be compensated by a field rescaling, the value of $\Vmin$ will be unchanged. 
To see this, simply note that the extrema of any function $f(x)$ are invariant under rescaling of the dependent variable: this rescaling just scales the $x$-axis (see Fig.~\ref{fig:rescaling}). Thus
the invariance of $\Vmin$ follows as a special case.

This property of the effective potential, that it depends on the scale where it is calculated, may seem unsettling. 
One might imagine that the $\ln\frac{\muY}{\muX}$ terms can be removed by invoking canonical normalization, perhaps by demanding that the kinetic
term in the effective action be reset to $\frac{1}{2}(\partial_\mu \phi)^2$ after including all the quantum corrections. One cannot just demand this, however.
First of all, in the effective action even the two-derivative kinetic terms have complicated field dependence 
\be
\Gamma[\phi] = \cdots+ \frac{1}{2} Z_1[\phi] (\partial_\mu \phi_i)( \partial_\mu \phi_i) + \frac{1}{2} Z_2[\phi] \frac{\phi_i\phi_j}{\phi^2} (\partial_\mu \phi_i)( \partial_\mu \phi_j) + 
Y_1[\phi] \frac{1}{\phi^4} \Big( (\partial_\mu \phi_i)( \partial_\mu \phi_i) \Big)^2 + \cdots
\cdots \label{ZandY}
\ee
The functionals $Z_1,Z_2, Y_1$ etc. can be calculated from prototype diagrams as a power expansion in external momentum $p^\mu$ ~\cite{Fraser:1984zb}. Explicit results at 1-loop
for $Z_1$ and $Z_2$ can be found in~\cite{Tye:1996au} and for $Y_1$ in $\phi^4$ theory in~\cite{Fraser:1984zb}.
So it is not clear how one would go about canonically normalizing $\phi$ apart from some byzantine implicit nonlinear field redefinition.

A more elementary objection to re-normalizing the fields is that we have already  renormalized them.
In $\msbar$ the field strength renormalization $Z_1$ is unambiguously fixed, so there is no remaining freedom. One could have chosen a different normalization convention, 
for example, an on-shell renormalization for which the residue of the propagator at the physical pole is set to 1 exactly.
However, since these are conventions, physical quantities must be independent of them. Thus we can work in $\msbar$. Indeed, as we have seen $\Vmin$ is independent of the normalization condition.

\subsection{RG improved effective potential \label{sec:RGimp}}
We have shown that the effective potential depends on the scale $\muY$ where it is calculated. We have also shown that when $\muY$ is close to the scale $\muX$ where the couplings satisfy Eq.~\eqref{lamerel}, then the value of $\Vmin$ is independent of $\muY$. Now suppose $\muY$ is not close to $\muX$, so that  $\ln\frac{\muY}{\muX}$ is not small. Then we cannot work in fixed-order perturbation theory. For example, in the standard model, one normally extracts the $\msbar$ couplings
through perturbative threshold calculations at a scale $\muY \sim 100$~GeV near the weak scale; the equivalent of $\muX$ is a scale near where $\lambda \sim 0$ which is many orders of magnitude higher
in energy; so, $\muX \gg \muY$. The minimum of the effective potential is found by solving the RGE for the effective potential to evolve it from $\muY$ to $\muX$. Thus the potential at $\mu=\muX$ includes
terms to all orders in perturbation theory whether it is written in terms of $g(\muY)$ or $g(\muX)$. This seems to limit the use of fixed order relations like Eq.~\eqref{lamerel}.

In order to combine the $\lambda \sim e^4$ scaling, which is required for gauge-invariance, with resummation, we can exploit the observation from Section~\ref{sec:discussion} that $\Vmin$ is independent
of rescaling $\phi$. In particular, we see that we do not have to evolve the entire effective potential from $\muY$ to $\muX$. Instead, we can just evolve the couplings, and then compute the fixed order
potential directly at $\muX$. The scale $\muY$ only enters for determining the boundary conditions of the renormalization equations. Since the $\beta$ functions are all gauge invariant,
gauge invariance of the potential at the minimum then automatically follows from the
arguments in Section~\ref{sec:GIphys}.

To be concrete, the resummed effective potential is often written as
\be
V[\phi] = \frac{1}{4!} e^{4\Gamma(\mu=\phi)} \left[ \lambda_{\text{eff}}^{(0)}(\mu = \phi) +  \lambda_{\text{eff}}^{(1)}(\mu = \phi) +    \lambda_{\text{eff}}^{(2)}(\mu = \phi) + \cdots \right] \phi^4 \label{Voldis}
\ee
with $\Gamma(\mu) = \int_{\muY}^\mu \gamma(\mu') \frac{d \mu'}{\mu'}$ and  $\frac{1}{4!}\lambda_{\text{eff}}^{(j)}(\mu)\phi^4$ the order $\hbar^j$ term in the fixed-order calculation of the effective potential. We are proposing instead that one should use Eq.~\eqref{Vnewis}:
\be
\Vmin =
 \VLO(\muX)
 +  \VNLO(\muX) 
 +   \VNNLO(\muX) 
+ v_1   \frac{d}{d\phi} \VNLO\Big|_{\phi=\muX}
 + \frac{1}{2} v_1^2  \frac{d^2}{d\phi^2} \VLO\Big|_{\phi=\muX} + \cdots \label{Vnewis2}
\ee
where $\VLO \sim \hbar$, $\VNLO \sim \hbar^2$, $v_1 \sim \hbar$ and so on, using $\lambda \sim\hbar$.
Using Eq.~\eqref{Vnewis2}, the value of $\Vmin$ will be gauge invariant order-by-order using this expansion, while $\Vmin$ computed using Eq.~\eqref{Voldis} will not be.

Note that this consistent prescription is not equivalent to using dressed fields where the field strength renormalization is absorbed into the field. Such a redefinition was advocated in~\cite{Frere:1974ia} (based on~\cite{Fischler:1974ue}) as removing all of the gauge-dependence in the effective potential. First of all, that redefinition does not work as advertised: even after absorbing the field strength renormalization, the effective potential is still gauge-dependent~\cite{Fukuda:1975di,Nielsen:1975fs}. Secondly, we are still using the same fields
and $\msbar$ normalizations relevant for $S$-matrix calculations, which greatly facilitates applications to the Standard Model.
Furthermore, since 
 $\VNLO(\muX)$ is  known in the Standard Model~\cite{Ford:1992pn}
 (and in general renormalizable theories~\cite{Martin:2001vx}) and 
 since  the daisy graphs begin at NNLO in Landau gauge,
Eq. \eqref{Vnewis} can immediately be implemented for many theories of direct phenomenological interest.  The effect of using Eq.~\eqref{Vnewis2} instead of Eq.~\eqref{Voldis} for the Standard Model effective potential is discussed in~\cite{us2}.

\section{Higher dimension operators \label{sec:HD}}
As we have seen, $\Vmin$ is gauge invariant but  the vev at the minimum, $v=\langle \phi \rangle$ is not.
 One reason one would like $v$ to be physical is that it gives a criteria for when higher dimension operators can have a significant effect. For example, 
one might expect an operator $\Delta \cL  =- \frac{1}{\Lambda^2} \phi^6$ to be relevant at field values where this term contributes of the same order as the leading one: 
$\Lambda \sim\frac{1}{\sqrt{\lambda(v)}} v$. Since $v$ is $\xi$-dependent, this criteria is not meaningful.

The correct way to evaluate the influence of a gauge-invariant higher-dimension operator is to add it to the classical potential and see how $\Vmin$ is affected. So let us add $-\frac{1}{\Lambda^2} \phi^6$ to the Lagrangian (that is, we add $+\frac{1}{\Lambda^2} \phi^6$ to the classical potential). First, consider
 the case when $\Lambda$ is very small (so this new term is large). Then the sensitive cancellation between $\lambda$ and $e^4$ will be  severely disrupted and the new quantum minimum will disappear. With the sign given, the only minimum would be at $\phi=0$ with $\Vmin=0$. If we flip the sign, then the potential will be unbounded from below and again there will be no quantum minimum, with $V=0$ still the only extremum and clearly gauge invariant.

Now suppose this new term has a very small effect. To be precise, recall that the quantum minimum in the absence of this term occurs at leading order at the scale $\phi = \muX$ where Eq.~\eqref{lamerel} is satisfied. Thus if this term is very small, at LO, it shifts the minimum to
\be
\Vmin = \muX^4
 \left[ - \frac{3}{8}    \frac{e^4 \hbar}{16 \pi^2}
  + \frac{\muX^2}{\Lambda^2} + \cO(\frac{\muX^4}{\Lambda^4} )\right]
  \label{Vminmod}
\ee
This is gauge invariant, but trivially so. 

 The key observation is that to compute subleading corrections when the effect is small, we cannot simply plug the next-to-leading order vev into
 $\frac{1}{\Lambda^2}\phi^6$. Doing so would give a gauge-dependent correction.  Instead, we must carefully re-evaluate our perturbation expansion in the presence of this new term.
 
If we are careful we must commit to whether this new term is important or not. If we say it is important, then minimally, the new term in Eq.~\eqref{Vminmod} must be comparable to the original one. That means that $\Lambda^2 \sim \muX^2 \frac{16\pi^2}{e^4 \hbar}$.  This power-counting means that the 1-loop minimum is not at the scale $\muX$ but rather at a new scale $\muZ$ where
\be
\lambda(\muZ) \equiv \frac{\hbar}{16\pi^2} e^4(\muZ)\Big[6-36 \ln [e(\muZ) ]\Big] 
-36 \frac{\muZ^2}{ \Lambda^2}
\label{lamerelZ}
\ee
which gives
\be
\Vmin = \muZ^4
 \left[- \frac{3}{8}    \frac{e^4 \hbar}{16 \pi^2}
  -\frac{1}{2} \frac{\muZ^2}{\Lambda^2} + \cO(e^6 )\right]
\ee
As before, the first nontrivial check will be at order $e^6$. To perform this check, one would need to compute the 2-loop graphs with this modification as well as the daisy resummation in the $R_\xi$ gauges. 

The main implication of the discussion in this section is that arguments about the relevance of higher dimension operators must be refined to have a consistent perturbative expansion. 
For example, one might traditionally calculate $\phi_{\text{min}}$ defined as the field value at the minimum in Landau gauge, and compare it to $\mpl$ to see gravitational effects on the effective potential. This comparison is not gauge invariant. Instead, one must actually insert the new operators and see how physical quantities determined by the effective potential change. An example implementation
of this procedure for the Standard Model is discussed in~\cite{us2}.

\section{Conclusions \label{sec:conc}}
In this paper we have resolved a long-standing puzzle about gauge dependence in the Coleman-Weinberg model. We have shown that the potential
at the minimum is gauge-invariant at the first non-trivial order in perturbation theory. The invariance
is only manifest if an appropriate power counting is used and if $\Vmin$ is expressed in terms of another gauge-invariant scale, such as a value of the $\msbar$ 
parameter $\mu$ where the couplings satisfy some relation. 
To establish gauge invariance 
required a calculation of  the full 2-loop Coleman-Weinberg potential in a general gauge to order $e^6$ as well as an infinite number of daisy diagrams which
are also relevant at next-to-leading order. 

Besides providing the first explicit check that a potential is gauge invariant at a loop-generated minimum, our work has implications for how effective
potentials should be used consistently. First of all, we showed that one cannot simply truncate to fixed order in the usual loop expansion. Some terms at a given
order must be dropped, and other terms from diagrams of arbitrarily high-loop order must be included. Only then will the potential have desired gauge-invariance properties. At leading order in the appropriate power counting, only a finite number of diagrams contribute. At next-to-leading order, the first infinite class of diagrams becomes relevant in a general gauge and must be resummed. A propitious feature of Landau gauge, $\xi=0$, is that this resummation is postponed to next-to-next-to-leading order, competitive with 3-loop graphs in the normal loop expansion. Nevertheless, even in Landau gauge at NLO, one {\it must} truncate the effective potential appropriately to have self-consistent results. This result is consistent with recent observations in~\cite{Martin:2014bca,Elias-Miro:2014pca}.

We also discussed how the consistent use we advocate is compatible with an RG-improved effective potential. The key is to  run the couplings first to a scale where the 1-loop potential, appropriately truncated, has its minimum. This scale is gauge-invariant.
 Then one can
add corrections to the effective potential around that scale order-by-order in perturbation theory using couplings at that scale. This essentially amounts to
dropping the field-strength renormalization contribution to the resummed effective potential. However, we are not performing any field redefinitions. Thus the same threshold corrections can be used to convert observables to $\msbar$ parameters at the low scale, and then run to the scale near the quantum minimum. 
The effective potential computed this way is still gauge-dependent. Moreover, even in Landau gauge (or any other gauge), the potential
is different if we first run the couplings and then calculate the potential or if we calculate the potential and then run it by solving its RGE. The difference
is not an artifact of the truncation in perturbation theory: the exact potentials computed these two ways would differ. However, this difference,
like the difference between potentials computed in different gauges, should not affect physical observables. We have shown that indeed it does not affect
the value of the potential at the minimum.

Our work definitively resolves the question of how one physical quantity, the true vacuum energy in a theory, can be computed consistently in perturbation theory.
We also know that $S$-matrix elements are gauge-invariant.
There are of course many other quantities that effective actions are used to calculate, and it would be very interesting to understand definitively how
they can be computed consistently as well. To compute the scalar
and vector masses, one would need to resum
the daisy contributions to the scalar and vector self-energy graphs
at small momentum, supplementing the results of~\cite{Kang:1974yj}. Presumably
these daisies would cancel the contribution from 2-loop effective potential,
leaving Kang's result intact. 
Another interesting quantity is the tunneling rate between two extrema.
A traditional tunneling rate calculation depends on unphysical regions of the potential~\cite{Espinosa:1995se,Isidori:2001bm}, away from the minima. However, 
any rate calculation should also depend on the kinetic terms in the effective action, so there are new opportunities for cancellation.
Although there has been some progress in understanding how tunneling rates may be consistent~\cite{Metaxas:1995ab,Strumia:1998nf}, there are, to our knowledge, no explicit demonstrations.

At finite temperature, the critical temperature for a phase transition should be physical. 
The spectrum of gravity waves as the universe cools through this transition should also be physical.
While some progress has been made on showing that these quantities can be computed in a gauge-invariant way in perturbation theory~\cite{Boyanovsky:1996dc,Patel:2011th,Wainwright:2011qy,Morrissey:2012db} the question does not seem to be completely resolved. Other apparently gauge-dependent quantities
include the compactification radius in certain extra-dimensional models~\cite{Kunstatter:1985yt,Nielsen:1987ht} and the inflation scale in 
Higgs-inflation models~\cite{Cook:2014dga}. 

Often the gauge-dependence in a quantity extracted from the effective potential originates in dependence on field values. For example, the instability
scale in the Standard Model $\Lambda_I$, defined as the value of $\phi$ where the potential goes negative~\cite{Buttazzo:2013uya}, is gauge-dependent~\cite{DiLuzio:2014bua}. The scale where the Standard Model effective potential is maximal, $\Lambda_{\text{max}}$, is also sometimes used
to make physical arguments~\cite{Hook:2014uia}, despite its gauge-dependence. A common criterion for stability in the Standard Model is that the value  $v$ of the field at the minimum
be less than $\mpl$. Since $v$ is gauge-dependent, so is this criterion~\cite{Loinaz:1997az}. Even if one asks only that the potential at the minimum, $\Vmin$ be less than $0$, which {\it is} a gauge-invariant criterion, one would presumably still want to know when this bound is affected by Planck-suppressed or
other higher-dimension operators. To this end, we have sketched one possible solution in Section~\ref{sec:HD}: one should add these new operators to the theory
and see how $\Vmin$ changes. Since the new operators affect the perturbation expansion, this procedure is not equivalent to replacing $\phi \to v$ in these operators
and seeing how the vacuum energy changes. With these examples, we are optimistic that apparent gauge-invariance of many physical quantities can be eliminated by the self-consistent use of an appropriate perturbation expansion.

\section{Acknowledgements}
\label{sec:acknowledgements}
We thank M. Reece, M. Ramsey-Musolf, M. Strassler and A. Strumia for helpful discussions.
The authors are supported in part by grant DE-SC003916 from the Department of Energy. AA is supported in part by the Stolt-Nielsen Fund for Education of the
the American-Scandinavian Foundation and the Norway-America Association. WF is supported in part by the Lord Rutherford Memorial Research Fellowship.

\bibliography{vsqed}

\bibliographystyle{utphys}

\end{document}